\documentclass{sig-alternate-05-2015}

\CopyrightYear{2016} 
\setcopyright{acmcopyright}
\conferenceinfo{SIGMOD'16,}{June 26-July 01, 2016, San Francisco, CA, USA}
\isbn{978-1-4503-3531-7/16/06}\acmPrice{\$15.00}
\doi{http://dx.doi.org/10.1145/2882903.2915213}

\usepackage{pgfplotstable}
\usepackage{booktabs} 
\usepackage{siunitx}
\usepackage{color}
\usepackage{tabularx}
\usepackage{multirow}
\usepackage{balance}
\usepackage{scrextend}
\usepackage{listings}
\usepackage{subcaption}
\usepackage{tikz}
\usetikzlibrary{positioning}
\usepackage{pifont}
\usepackage{url,xspace}
\usepackage{algorithm}
\usepackage{amsmath}
\usepackage{bigstrut,cite}
\usepackage{float}
\usepackage[bookmarks=false]{hyperref}
\usepackage{cleveref}
\usepackage{bold-extra}

\usetikzlibrary{shapes,arrows}

\newcolumntype{P}[1]{>{\centering\arraybackslash}p{#1}}

\newcommand{\R}{\mathbb{R}}
\newcommand{\EH}{EmptyHeaded\xspace} 
\newcommand{\EHSMALL}{EH\xspace} 
\newcommand{\out}{\textsc{out}}
\newcommand{\TheGame}{{\tt hybrid}}

\newcommand{\card}[1]{\ensuremath{\left\vert#1\right\vert}}

\newcommand{\blank}[1]{\hspace*{#1}\linebreak[0]}

\newcommand{\uint}{\texttt{uint}\xspace}
\newcommand{\uints}{\texttt{uints}\xspace}
\newcommand{\bitset}{\texttt{bitset}\xspace}
\newcommand{\bitsets}{\texttt{bitsets}\xspace}
\newcommand{\pshort}{\texttt{pshort}\xspace}

\newcommand{\variant}{\texttt{variant}\xspace}
\newcommand{\bitpacked}{\texttt{bitpacked}\xspace}

\newcommand{\andand}{\texttt{and}\xspace}

\def\compactify{\itemsep=0pt \topsep=0pt \partopsep=0pt \parsep=0pt}

\definecolor{linenumbercolor}{rgb}{0.5,0.5,0.5}
\newdef{definition}{Definition}
\newtheorem{example}{Example}[section]
\newfloat{algorithm}{t}{lop}

\title{\EH: A Relational Engine for Graph Processing}
\numberofauthors{4}
\author{
\alignauthor Christopher R. Aberger\\
\affaddr{Stanford University}\\
  \email{caberger@stanford.edu}
\alignauthor Susan Tu \\
\affaddr{Stanford University}\\
  \email{sctu@stanford.edu}
\alignauthor Kunle Olukotun \\
\affaddr{Stanford University}\\
  \email{kunle@stanford.edu}
\and
\alignauthor Christopher R\'e\\
\affaddr{Stanford University}\\
  \email{chrismre@cs.stanford.edu}
}

\begin{document}
\pagestyle{empty}

\maketitle

\begin{abstract}
There are two types of high-performance graph processing engines: low- and high-level engines. Low-level engines (Galois, PowerGraph, Snap) provide optimized data structures and computation models but require users to write low-level imperative code, hence ensuring that efficiency is the burden of the user. In high-level engines, users write in query languages like datalog (SociaLite) or SQL (Grail). High-level engines are easier to use but are orders of magnitude slower than the low-level graph engines. We present \EH, a high-level engine that supports a rich datalog-like query language and achieves performance comparable to that of low-level engines. At the core of \EH's design is a new class of join algorithms that satisfy strong theoretical guarantees but have thus far not achieved  performance comparable to that of specialized graph processing engines. To achieve high performance, \EH introduces a new join engine architecture, including a novel query optimizer and data layouts that leverage single-instruction multiple data (SIMD) parallelism. With this architecture, \EH outperforms high-level approaches by up to three orders of magnitude on graph pattern queries, PageRank, and Single-Source Shortest Paths (SSSP) and is an order of magnitude faster than many low-level baselines. We validate that \EH competes with the best-of-breed low-level engine (Galois), achieving comparable performance on PageRank and at most 3x worse performance on SSSP.
\end{abstract}

\begin{category}
{H.2}{Information Systems}{Database Management System Engines}
\end{category}


\begin{figure*}
  \centering
  \includegraphics[width=\linewidth,height=3cm]{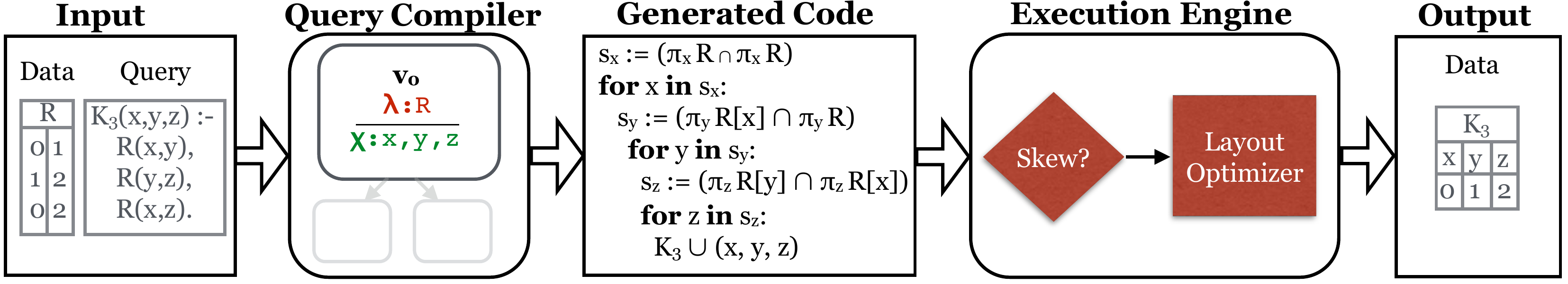}
  \caption{The \EH{} engine works in three phases:
(1) the query compiler translates a high-level datalog-like query
into a logical query plan represented as a
GHD (a hypertree with a single node here), replacing
the traditional role of relational algebra; (2) code is generated for the execution engine by translating the GHD into a
series of set intersections and loops; and (3) the execution engine
performs automatic algorithmic and layout decisions based upon
skew in the data.}
  \label{fig:system}
\end{figure*}

\vspace{8mm}
\section{Introduction}

The massive growth in the volume of graph data from social and biological networks has created a need for efficient graph processing engines. As a result, there has been a flurry of activity around designing specialized graph analytics engines \cite{Hong:2012:GDE:2150976.2151013,Gonzalez:2012:PDG:2387880.2387883,Nguyen:2013:LIG:2517349.2522739,snap,Shun:2013:LLG:2517327.2442530}. These specialized engines offer new programming models that are either (1) low-level, requiring users to write code imperatively or (2) high-level, incurring large performance gaps relative to the low-level approaches. In this work, we explore whether we can meet the performance of low-level engines while supporting a high-level relational (SQL-like) programming interface.

Low-level graph engines outperform traditional relational data processing engines on common benchmarks due to (1) asymptotically faster algorithms~\cite{Schank:2005:FCL:2154763.2154820,ngo2012worst} and (2) optimized data layouts that provide large constant factor runtime improvements~\cite{Hong:2012:GDE:2150976.2151013}. We describe each point in detail:

\begin{enumerate} \compactify
  \item Low-level graph engines \cite{Hong:2012:GDE:2150976.2151013,Gonzalez:2012:PDG:2387880.2387883,Nguyen:2013:LIG:2517349.2522739,snap,Shun:2013:LLG:2517327.2442530} provide iterators and domain-specific primitives, with which users can write asymptotically faster algorithms than what traditional databases or high-level approaches can provide. However, it is the burden of the user to write the query properly, which may require system-specific optimizations. Therefore, optimal algorithmic runtimes can only be achieved through the user in these low-level engines.  
\item Low-level graph engines use optimized data layouts to efficiently manage the sparse relationships common in graph data. For example, optimized sparse matrix layouts are often used to represent the edgelist relation \cite{Hong:2012:GDE:2150976.2151013}. High-level graph engines also use sparse layouts like tail-nested tables \cite{seo2013socialite} to cope with sparsity.
\end{enumerate}

Extending the relational interface to match these guarantees is
challenging. While some have argued that traditional engines can be
modified in straightforward ways to accommodate graph
workloads~\cite{fancase,aref2015design}, order of magnitude
performance gaps remain between this approach and low-level engines
\cite{snap,Nguyen:2013:LIG:2517349.2522739,seo2013socialite}. Theoretically,
traditional join engines face a losing battle, as all pairwise join engines are {\em provably suboptimal} on many common graph
queries~\cite{ngo2012worst}. For example, low-level specialized
engines execute the ``triangle listing'' query, which is common in
graph workloads \cite{milo2002network,newman2003structure}, in time
$O(N^{3/2})$ where $N$ is the number of edges in the graph. Any pairwise relational algebra plan takes at least
$\Omega(N^2)$, which is asymptotically worse than the specialized engines by a factor
of $\sqrt{N}$. This asymptotic suboptimality is often inherited by
high-level graph engines, as there has not been a general way to
compile these queries that obtains the correct
asymptotic bound~\cite{seo2013socialite,fancase}. Recently, new multiway
join algorithms were discovered that obtain the correct asymptotic
bound for any graph pattern or join~\cite{ngo2012worst}.

These new multiway join algorithms are by themselves not enough to
close the gap. LogicBlox~\cite{aref2015design} uses multiway join
algorithms and has demonstrated that they can support a rich set of
applications. However, LogicBlox's current engine can be orders of
magnitude slower than the specialized engines on graph
benchmarks (see \Cref{sec:experiments}). This leaves open the question
of whether these multiway joins are destined to be slower than
specialized approaches.

We argue that an engine based on multiway join algorithms can close this gap, but it requires a novel architecture (\Cref{fig:system}), which forms our main contribution. Our architecture includes a novel query compiler based on \emph{generalized hypertree decompositions} (GHDs) ~\cite{gottlob2005hypertree,chekuri1997conjunctive} and an execution engine designed to exploit the low-level layouts necessary to increase single-instruction multiple data (SIMD) parallelism. We argue that these techniques demonstrate that multiway join engines can compete with low-level graph engines, as our prototype is faster than all tested engines on graph pattern queries (in some cases by orders of magnitude) and competitive on other common graph benchmarks.

We design \EH around tight theoretical guarantees and data layouts optimized for SIMD parallelism.

\paragraph*{GHDs as Query Plans} 
The classical approach to query planning uses relational algebra, which facilitates optimizations such as early aggregation, pushing down selections, and pushing down projections. In \EH, we need a similar framework that supports multiway (instead of pairwise) joins. To accomplish this, based off of an initial prototype developed in our group \cite{tu2015duncecap}, we use \emph{generalized hypertree decompositions} (GHDs)~\cite{gottlob2005hypertree} for logical query plans in \EH. GHDs allow one to apply the above classical optimizations to multiway joins. GHDs also have additional bookkeeping information that allow us to bound the size of intermediate results (optimally in the worst case). These bounds allow us to provide asymptotically stronger runtime guarantees than previous worst-case optimal join algorithms that do not use GHDs (including LogicBlox).\footnote{LogicBlox has described a (non-public) prototype with an optimizer similar but distinct from GHDs. With these modifications, LogicBlox's relative performance improves similarly to our own. It, however, remains at least an order of magnitude slower than \EH.} As these bounds depend on the data and the query it is difficult to expect users to write these algorithms in a low-level framework. Our contribution is the design of a novel query optimizer and code generator based on GHDs that is able to achieve the above results via a high-level query language.

\paragraph*{Exploiting SIMD: The Battle With Skew}
Optimizing relational databases for the SIMD hardware trend has become an increasingly hot research topic \cite{Li:2013:BFS:2463676.2465322,Raman:2013:DBA:2536222.2536233,Zhou:2002:IDO:564691.564709}, as the available SIMD parallelism has been doubling consistently in each processor generation.\footnote{The Intel Ivy Bridge architecture, which we use in this paper, has a SIMD register width of 256 bits. The next generation, the Intel Skylake architecture, has 512-bit registers and a larger number of such registers.} Inspired by this, we exploit the link between SIMD parallelism and worst-case optimal joins for the first time in \EH.  Our initial prototype revealed that during query execution, unoptimized set intersections often account for 95\% of the overall runtime in the generic worst-case optimal join algorithm. Thus, it is critically important to optimize set intersections and the associated data layout to be well-suited for SIMD parallelism. This is a challenging task as graph data is highly skewed, causing the runtime characteristics of set intersections to be highly varied. We explore several sophisticated (and not so sophisticated) layouts and algorithms to opportunistically increase the amount of available SIMD parallelism in the set intersection operation. Our contribution here is an automated optimizer that, all told, increases performance by up to three orders of magnitude by selecting amongst multiple data layouts and set intersection algorithms that use skew to increase the amount of available SIMD parallelism.

We choose to evaluate \EH on graph pattern matching queries since pattern queries are naturally (and classically) expressed as join queries. We also evaluate \EH on other common graph workloads including PageRank and Single-Source Shortest Paths (SSSP). We show that \EH consistently outperforms the standard baselines~\cite{fancase} by 2-4x on PageRank and is at most 3x slower than the highly tuned implementation of Galois \cite{Nguyen:2013:LIG:2517349.2522739} on SSSP. However, in our high-level language these queries are expressed in 1-2 lines, while they are over 150 lines of code in Galois. For reference, a hand-coded C implementation with similar performance to Galois is 1000 lines.

\paragraph*{Contribution Summary} This paper introduces the EmptyHeaded engine and demonstrates that a novel architecture can enable multi-way join engines to compete with specialized low-level graph processing engines. We demonstrate that \EH outperforms specialized engines on graph pattern queries while remaining competitive on other workloads. To validate our claims we provide comparisons on standard graph benchmark queries that the specialized engines are designed to process efficiently.

A summary of our contributions and an outline is as follows:

\begin{itemize} \compactify
  \item We describe the first worst-case optimal join processing engine
  to use GHDs for logical query plans. We describe how GHDs enable
  us to provide a tighter theoretical guarantee than previous worst-case
  optimal join engines (\Cref{sec:qc}). Next, we validate that the optimizations GHDs enable provide more than a three orders of magnitude performance advantage over previous worst-case optimal query plans (\Cref{sec:experiments}).
  \item We describe the architecture of the first worst-case optimal 
  execution engine that optimizes for skew at several levels of granularity within
  the data. We present a series of automatic optimizers
  to select intersection algorithms and set layouts
  based on data characteristics at runtime (\Cref{sec:execution_engine}). We demonstrate that our automatic optimizers can result in up to a three orders of magnitude performance improvement on common graph pattern queries (\Cref{sec:experiments}).
  \item We validate that our general purpose engine can compete with specialized
  engines on standard benchmarks in the graph domain (\Cref{sec:experiments}).
  We demonstrate that on cyclic graph pattern queries our approach outperforms graph engines by 2-60x and LogicBlox by three orders of magnitude. We demonstrate on PageRank and Single-Source Shortest Paths that our approach remains competitive, at most 3x off the highly tuned Galois engine (\Cref{sec:experiments}).
\end{itemize}

\section{Preliminaries}

We briefly review the worst-case optimal join algorithm, trie data structure, and query language at the core of the \EH design. The worst-case optimal join algorithm, trie data structure, and query language presented here serve as building blocks for the remainder of the paper.

\label{sec:backround}
\subsection{Worst-Case Optimal Join Algorithms}
\label{sec:wc_back}
We briefly review worst-case optimal join algorithms, which are used in
\EH. We present these results informally and refer the reader to Ngo et al.~\cite{worst}
for a complete survey. The main idea is that one can place (tight) bounds on
the maximum possible number of tuples returned by a query and then develop algorithms whose runtime guarantees match these worst-case bounds.
For the moment, we consider only join queries (no
projection or aggregation), returning to these richer queries in \Cref{sec:qc}.

A {\bf {\em hypergraph}} is a pair $H = (V, E)$, consisting of a
nonempty set $V$ of vertices, and a set $E$ of subsets of $V$, the
hyperedges of $H$. Natural join queries and graph pattern queries can
be expressed as hypergraphs \cite{gottlob2005hypertree}. In particular, there is
a direct correspondence between a query and its hypergraph: there is a
vertex for each attribute of the query and a hyperedge for each
relation. We will go freely back and forth between the query and the
hypergraph that represents it.

A recent result of Atserias, Grohe, and Marx~\cite{agm} (AGM) showed
how to tightly bound the worst-case size of a join query using a
notion called a fractional cover. Fix a hypergraph $H=(V,E)$. Let $x
\in \R^{|E|}$ be a vector indexed by edges, i.e., with one component
for each edge, such that $x \geq 0$; $x$ is a {\em
  feasible cover} (or simply feasible) for $H$ if
\[ \text{ for each } v \in V \text{ we have } \sum_{e \in E : e \ni v} x_{e} \geq 1 \]
A feasible cover $x$ is also called a {\em fractional hypergraph cover}
in the literature. AGM showed that if $x$ is feasible then it forms an upper bound of the
query result size $|\out|$ as follows:
\begin{equation}
  |\out| \leq \prod_{e \in E} |R_{e}|^{x_e}
\label{eq:size:bound}\end{equation}
For a query $Q$, we denote $\mathsf{AGM}(Q)$ as the smallest such
right-hand side.\footnote{One can find the best bound,
  $\mathsf{AGM}(Q)$, in polynomial time: take the $\log$ of
  Eq.~\ref{eq:size:bound} and solve the linear program.}

\begin{example}
  For simplicity, let $|R_e| = N$ for $e \in E$. Consider the triangle
  query, $R(x,y) \bowtie S(y,z) \bowtie T(x,z)$,  a feasible cover is
  $x_R = x_S = 1$ and $x_{T} = 0$. Via
  Equation~\ref{eq:size:bound}, we know that $|\out| \leq
  N^{2}$. That is, with $N$ tuples in each relation we cannot produce
  a set of output tuples that contains more than $N^2$. However, a
  tighter bound can be obtained using a different fractional cover $x
  = (\frac{1}{2},\frac{1}{2},\frac{1}{2})$.
  Equation~\ref{eq:size:bound} yields the upper bound $N^{3/2}$. Remarkably, this
  bound is tight if one considers the complete graph on $\sqrt{N}$
  vertexes. For this graph, this query produces $\Omega(N^{3/2})$ tuples, which shows that the optimal solution can be tight up to constant factors.
\end{example}

The first algorithm to have a running time matching these worst-case size bounds is the NPRR algorithm \cite{ngo2012worst}. 
An important property for the set intersections in the NPRR algorithm is what we call the 
{\em min property}: the running time of the intersection
algorithm is upper bounded by the length of the {\em smaller} of the
two input sets. When the min property holds, a worst-case optimal
running time for {\em any} join query is guaranteed. In fact, for {\em any}
join query, its execution time can be upper bounded by
$\mathsf{AGM}(Q)$. A simplified high-level description of the algorithm is presented in \Cref{fig:worst_case}. It was also shown that any pairwise join plan must be slower by asymptotic factors. However, we show in \Cref{sec:qp_ghd} that these optimality guarantees can be improved for non-worst-case data or more complex queries. 

\begin{algorithm}
  \begin{lstlisting}[
        basicstyle = \small,
        language = C,
        numbers = left,
        mathescape]
  //Input: Hypergraph $H = (V,E)$, and a tuple $t$.
  Generic-Join($V$,$E$,$t$):
    if $|V| = 1$ then return $\cap_{e \in E} R_e[t]$.
    Let $I = \{ v_{1} \}$ // the first attribute.
    Q $\leftarrow \emptyset$ // the return value
    // Intersect all relations that contain $v_{1}$
    // Only those tuples that agree with $t$.
    for every $t_v \in \cap_{e \in E : e \ni v_{1}} \pi_{I} (R_e[t])$ do
      $Q_t$ $\leftarrow$ Generic-Join($V - I$, $E$, $t :: t_v$ )
      Q $\leftarrow Q \cup \{ t_v \} \times Q_t $
    return Q
  \end{lstlisting}
  \caption{Generic Worst-Case Optimal Join Algorithm}
  \label{fig:worst_case}
\end{algorithm}

\begin{table*}
  \small
  \centering
  \begin{tabular}{ll}
  \toprule
  Name & Query Syntax \\
  \midrule
  Triangle &
   \begin{lstlisting}[
    language=C,
    basicstyle=\ttfamily,
    keywordstyle=\bfseries,
    showstringspaces=false]
Triangle(x,y,z) :- R(x,y),S(y,z),T(x,z).
  \end{lstlisting} \\

  4-Clique &
   \begin{lstlisting}[
    language=C,
    basicstyle=\ttfamily,
    keywordstyle=\bfseries,
    showstringspaces=false]
4Clique(x,y,z,w) :- R(x,y),S(y,z),T(x,z),U(x,w),V(y,w),Q(z,w).
  \end{lstlisting} \\

  Lollipop & 
  \begin{lstlisting}[
    language=C,
    basicstyle=\ttfamily,
    keywordstyle=\bfseries,
    showstringspaces=false]
Lollipop(x,y,z,w) :- R(x,y),S(y,z),T(x,z),U(x,w).
  \end{lstlisting} \\

  Barbell & 
  \begin{lstlisting}[
    language=C,
    basicstyle=\ttfamily,
    keywordstyle=\bfseries,
    showstringspaces=false]
Barbell(x,y,z,x',y',z') :- R(x,y),S(y,z),T(x,z),U(x,x'),R'(x',y'),S'(y',z'),T'(x',z').
  \end{lstlisting} \\
\hline
  Count Triangle &  \begin{lstlisting}[
    language=C,
    basicstyle=\ttfamily,
    keywordstyle=\bfseries,
    showstringspaces=false]
CountTriangle(;w:long) :- R(x,y),S(x,z),T(x,z); w=<<COUNT(*)>>.
  \end{lstlisting} \\
\hline
  \multirow{1}{*}{PageRank} &  \begin{lstlisting}[
    language=C,
    basicstyle=\ttfamily,
    keywordstyle=\bfseries,
    showstringspaces=false]
N(;w:int) :- Edge(x,y); w=<<COUNT(x)>>.
PageRank(x;y:float) :- Edge(x,z); y= 1/N.
PageRank(x;y:float)*[i=5] :- Edge(x,z),PageRank(z),InvDeg(z); y=0.15+0.85*<<SUM(z)>>.
  \end{lstlisting} \\

  \multirow{1}{*}{SSSP} &  \begin{lstlisting}[
    language=C,
    basicstyle=\ttfamily,
    keywordstyle=\bfseries,
    showstringspaces=false]
SSSP(x;y:int) :- Edge("start",x); y=1.
SSSP(x;y:int)* :- Edge(w,x),SSSP(w); y=<<MIN(w)>>+1.
  \end{lstlisting} \\

  \bottomrule
  \end{tabular}
  \caption{Example Queries in \EH}
  \label{lst:eh_query_language}
\end{table*}

\subsection{Input Data}

\label{sec:input_data}

\EH{} stores all relations (input and output) in tries, which are multi-level data
structures common in column stores and graph engines \cite{stonebraker2005c,Hong:2012:GDE:2150976.2151013}.  

\paragraph*{Trie Annotations} The sets of values in the trie can optionally be associated with data values (1-1 mapping) that are used in aggregations. We call these associated values \emph{annotations} \cite{green2007provenance}. For example, a two-level trie annotated with a float value represents a sparse matrix or graph with edge properties. We show in \Cref{sec:experiments} that the trie data structure works well on a wide variety of graph workloads. 

\paragraph*{Dictionary Encoding} The tries in \EH{} currently support sets containing 32-bit values. 
 As is standard \cite{Raman:2013:DBA:2536222.2536233,Gonzalez:2012:PDG:2387880.2387883}, we use the popular database technique of dictionary encoding to build a \EH trie from input tables of arbitrary types. Dictionary encoding maps original data values to keys of another type---in our case 32-bit unsigned integers. The order of dictionary ID assignment affects the density of the sets in the trie, and as others have shown this can have a dramatic impact on overall performance on certain queries. Like others, we find that node ordering is powerful when coupled with pruning half the edges in an undirected graph \cite{Schank:2005:FCL:2154763.2154820}. This creates up to 3x performance difference on symmetric pattern queries like the triangle query. Unfortunately this optimization is brittle, as the necessary symmetrical properties break with even a simple selection. On more general queries we find that node ordering typically has less than a 10\% overall performance impact. We explore the effect of various node orderings in \Cref{sec:node_ordering}.

\begin{figure}
  \centering
  \includegraphics[width=0.5\textwidth,height=4.5cm]{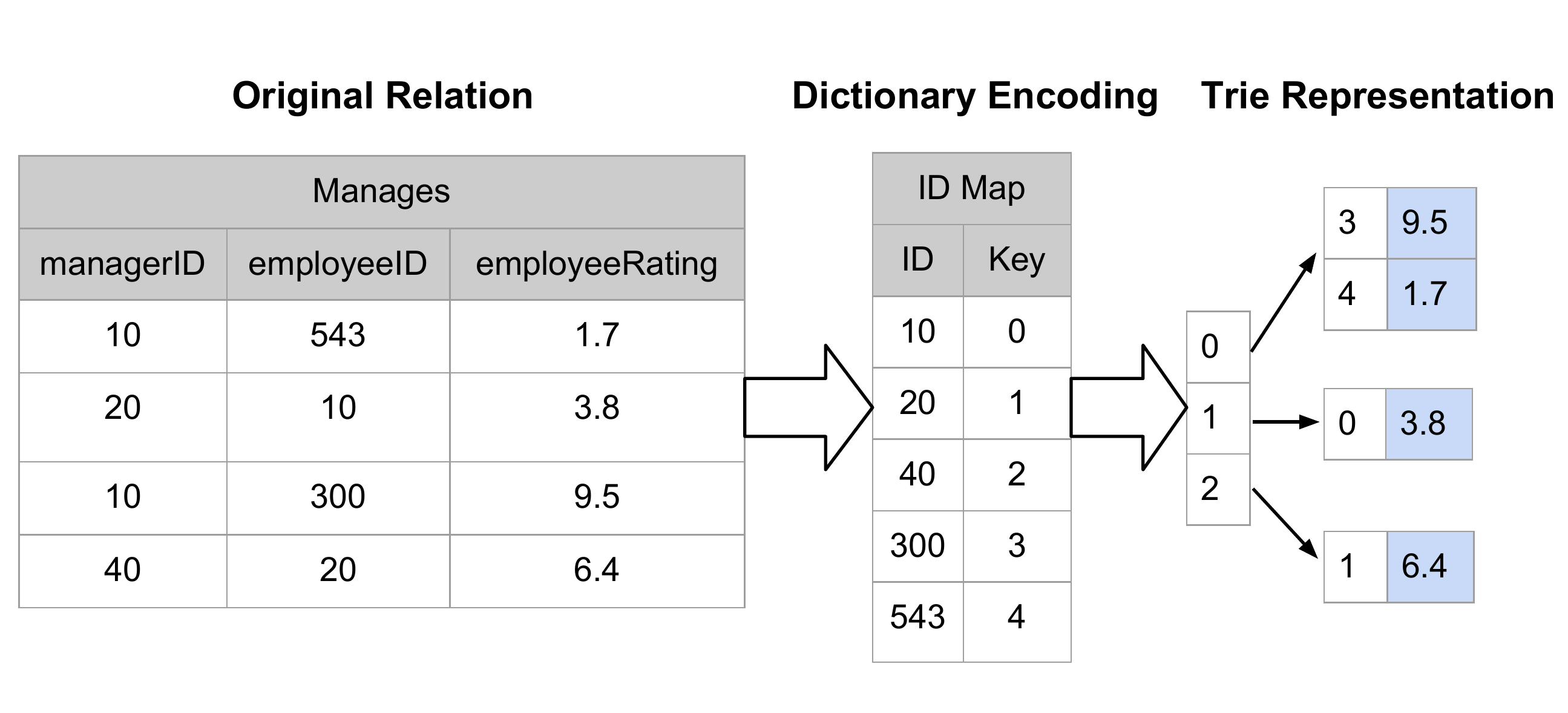}
  \centering 
  \caption{\EH{} transformations from a table to trie representation using attribute order (\emph{managerID,employerID}) and \emph{employerID} attribute annotated with \emph{employeeRating}.} 
  \label{fig:table_transform}
\end{figure}

 \paragraph*{Column (Index) Order} After dictionary encoding, our 32-bit value relations are next grouped into sets of distinct values based on their parent attribute (or column). We are free to select which level corresponds to each attribute (or column) of an input relation. As with most graph engines, we simply store both orders for each edge relation. In general, we choose the order of the attributes for the trie based on a global attribute order, which is analogous to selecting a single index over the relation. The trie construction process produces tries where the sets of data values can be extremely dense, extremely sparse, or anywhere in between. Optimizing the layout of these sets based upon their data characteristics is the focus of \Cref{sec:execution_engine}. 
The complete transformation process from a standard relational table to the trie representation in \EH{} is detailed in \Cref{fig:table_transform}. 

\subsection{Query Language}
Our query language is inspired by datalog and supports conjunctive
queries with aggregations and simple recursion (similar to LogicBlox
and SociaLite). In this section, we describe the core syntax for our
queries, which is sufficient to express the standard benchmarks we run
in \Cref{sec:experiments}. \Cref{lst:eh_query_language} shows the
example queries used in this paper. Above the first horizontal
line are conjunctive queries that express joins,
projections, and selections in the standard way \cite{cgqdl}. Our language has two
non-standard extensions: aggregations and a limited form of recursion. We overview both extensions next and provide an example in \Cref{sec:ext_query_language}.

\begin{figure*}
\begin{subfigure}[b]{0.30\linewidth} 
  \centering
  \includegraphics[height=3cm,width=4cm]{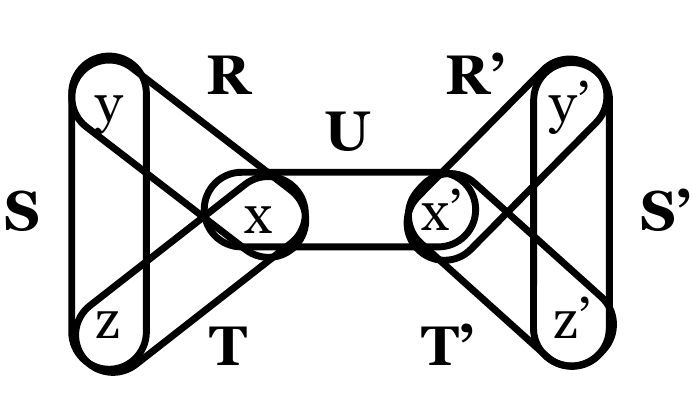}
  \caption{Hypergraph}
  \label{fig:barbell_hypergraph}
\end{subfigure}
\begin{subfigure}[b]{0.30\linewidth} 
  \centering
  \includegraphics[height=2.5cm,width=3cm]{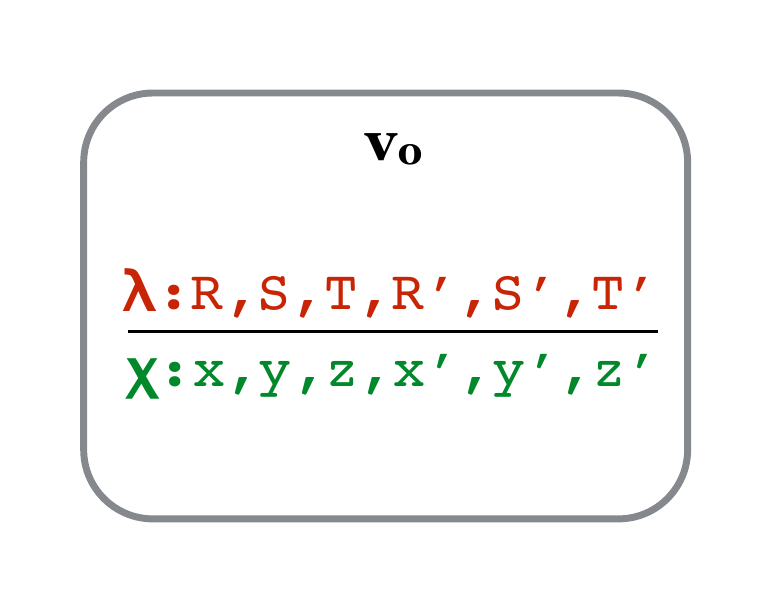}
  \caption{LogicBlox GHD}
  \label{fig:barbell_lb}
\end{subfigure}
\begin{subfigure}[b]{0.30\linewidth} 
  \centering
  \includegraphics[height=3cm,width=4.0cm]{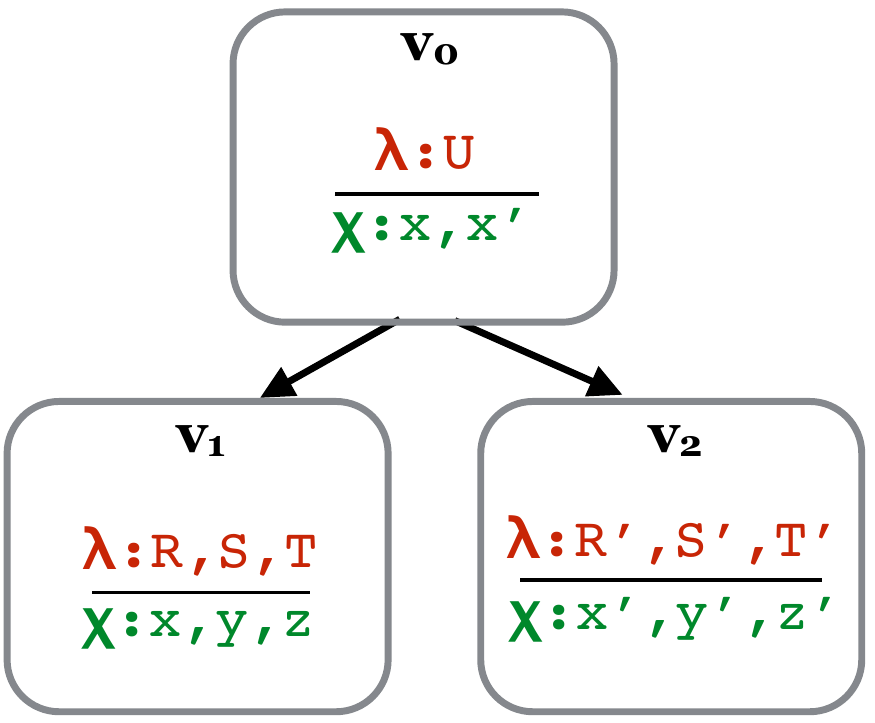}
  \caption{\EH{} GHD}
  \label{fig:barbell_ghd}
\end{subfigure}
  \centering 
  \caption{We show the Barbell query hypergraph and two possible GHDs for the query. A node $v$ in a GHD captures which relations should be joined with $\lambda(v)$ and which attributes should be retained with projection with $\chi(v)$. }
  \label{fig:ghd-lp}
\end{figure*}

\paragraph*{Aggregation}
Following Green et al. \cite{green2007provenance}, tuples can be
annotated in \EH, and these annotations support aggregations from
any semiring (a generalization of natural numbers equipped with a
notion of addition and multiplication). This enables \EH to support
classic aggregations such as \texttt{SUM}, \texttt{MIN}, or \texttt{COUNT}, but also
more sophisticated operations such as matrix multiplication. To specify the annotation, one uses a semicolon in the head
of the rule, e.g., \texttt{q(x,y;z:int)} specifies that each \texttt{x,y} pair will
be associated with an integer value with alias \texttt{z} similar to a \texttt{GROUP
BY} in SQL. In addition, the user expresses the aggregation operation
in the body of the rule. The user can specify an initialization value as any expression over the tuples' values and constants, while common
aggregates have default values. Directly below the first line in \Cref{lst:eh_query_language}, a typical
triangle counting query is shown.

\paragraph*{Recursion}
\EH supports a simplified form of recursion similar to Kleene-star or
transitive closure. Given an intensional or extensional relation $R$,
one can write a Kleene-star rule like:
\[ \textrm{R*}(\bar x) \quad \textrm{:-}\quad q(\bar x, \bar y) \]
The rule $R*$ iteratively applies $q$ to the current instantiation of $R$ to
generate new tuples which are added to $R$. It performs this iteration until
(a) the relation doesn't change (a fixpoint semantic) or (b) a user-defined convergence criterion is satisfied (e.g. a number of
iterations, \texttt{i=5}). Examples that capture familiar PageRank and Single-Source
Shortest Paths are below the second horizontal line in
\cref{lst:eh_query_language}.

\section{Query Compiler}

We now present an overview of the query compiler in \EH, which is the first worst-case optimal query compiler to enable early aggregation through its use of GHDs for logical query plans. We first discuss GHDs and their theoretical advantages. Next, we describe how we develop a simple optimizer to select a GHD (and therefore a query plan). Finally, we show how \EH translates a GHD into a series of loops, aggregations, and set intersections using the generic worst-case optimal join algorithm \cite{ngo2012worst}. Our contribution here is the design of a novel query compiler that provides tighter runtime guarantees than existing approaches. 

\label{sec:qc}

\subsection{Query Plans using GHDs}
\label{sec:qp_ghd}

As in a classical database, \EH needs an analog of relational algebra
to represent logical query plans. In contrast to traditional relational
algebra, \EH has multiway join operators. A natural approach would be
simply to extend relational algebra with a multiway join
algorithm. Instead, we advocate replacing relational algebra with GHDs, which allow us to make
non-trivial estimates on the cardinality of intermediate results. This enables optimizations, like early aggregation in \EH, that can be asymptotically faster than existing worst-case optimal engines. We first describe the motivation for using GHDs while formally describing their advantages next. 

\subsubsection{Motivation}

  A GHD is a tree similar to the abstract syntax tree of a relational algebra expression: nodes represent a join and projection operation, and edges indicate data dependencies. A node $v$ in a GHD captures which attributes should be retained (projection with $\chi(v)$) and which relations should be joined (with $\lambda(v)$). We consider all possible query plans (and therefore all valid GHDs), selecting the one where the sum of each node's runtime is the lowest. Given a query, there are many valid GHDs that capture the query. Finding the lowest-cost GHD is one goal of our optimizer.

Before giving the formal definition, we illustrate GHDs and their advantages by example:

\begin{example}
	\label{ex:barbell}

  \Cref{fig:barbell_hypergraph} shows a hypergraph of the Barbell
  query introduced in \Cref{lst:eh_query_language}. This query finds all pairs of triangles connected by a path of length one. Let $\out$ be the
  size of the output data. From our definition in
    \Cref{sec:wc_back}, one can check that the Barbell query has a
  feasible cover of $(\frac{1}{2}, \frac{1}{2}, \frac{1}{2}, 0,
  \frac{1}{2}, \frac{1}{2}, \frac{1}{2})$ with cost $6\times\frac{1}{2} = 3$
  and so runs in time $O(N^{3})$. In fact, this bound is worst-case
  optimal because there are instances that return $\Omega(N^3)$
  tuples. However, the size of the output ${\out}$ could be much
  smaller.

  There are multiple GHDs for the Barbell query. The
  simplest GHD for this query (and in fact for all queries) is a GHD
  with a single node containing all relations; the single node
  GHD for the Barbell query is shown in \Cref{fig:barbell_lb}. One
  can view all of LogicBlox's current query plans as a single
  node GHD. The single node GHD always represents a query plan which
  uses only the generic worst-case optimal join algorithm and
  no GHD optimizations. For the Barbell query, ${\out}$ is
  $N^{3}$ in the worst-case for the single node GHD.

  Consider the alternative GHD shown in \Cref{fig:barbell_ghd}.  This
  GHD corresponds to the following alternate strategy to the above
    plan: first list each triangle independently using the
    generic worst-case optimal algorithm, say on the vertices
  $(x,y,z)$ and then $(x',y',z')$. There are at most $O(N^{3/2})$
  triangles in each of these sets and so it takes only this time. Now,
  for each $(x,x') \in U$ we output all the triangles that contain $x$
  or $x'$ in the appropriate position. This approach is able to run in time $O(N^{3/2} +
  {\out})$ and essentially performs early aggregation if possible. This approach can be substantially faster when ${\out}$ is
  smaller than $N^3$. For example, in an aggregation query ${\out}$ is just a single scalar, and so the difference in runtime between the two GHDs can be quadratic in the size of the database. We describe
  how we execute this query plan in \Cref{sec:qp_codegen}. This type of optimization is currently not available in the
  LogicBlox engine.
  
  \label{ex:ghd:fast}\end{example}

\subsubsection{Formal Description}

We describe GHDs and their advantages formally next. 

\begin{definition}
 Let
$H$ be a hypergraph. A {\bf {\em generalized hypertree decomposition
    (GHD)}} of $H$ is a triple $D=(T, \chi, \lambda)$, where:
\begin{itemize}\compactify
\item  $T(V(T), E(T))$ is a tree;
\item $\chi: V (T) \rightarrow 2^{V(H)}$ is a function associating a set of vertices $\chi(v) \subseteq V(H)$ to each node $v$ of $T$;
\item $\lambda: V(T) \rightarrow 2^{E(H)}$ is a function associating a set of hyperedges to each vertex $v$ of $T$;
\end{itemize}

such that the following properties hold:
\begin{itemize}
\item[1.] For each $e \in E(H)$, there is a node $v \in V(T)$ such that $e \subseteq \chi(v)$ and $e \in \lambda(v)$. 
\item[2.] For each $t \in V(H)$, the set $\{v \in V(T) | t \in \chi(v)\}$ is connected in $T$.
\item[3.] For every $v \in V(T)$, $\chi(v) \subseteq \cup
  \lambda(v)$.
\end{itemize}
\end{definition}

A GHD can be thought of as a labeled (hyper)tree, as illustrated in
\Cref{fig:ghd-lp}. Each node of the tree $v$ is labeled; $\chi(v)$
describes which attributes are ``returned'' by the node $v$--this
exactly captures projection in traditional relational algebra. The
label $\lambda(v)$ captures the set of relations that are present in a
(multiway) join at this particular node. The first property says that
every edge is mapped to some node, and the second property is the
famous {\it ``running intersection
  property''}~\cite{abiteboul1995foundations} that says any attribute
must form a connected subtree. The third
  property is redundant for us, as any GHD violating
  this condition is not considered (has infinite width which we describe next). 

Using GHDs, we can define a non-trivial cardinality estimate based on
the sizes of the relations. For a node $v$, define $Q_v$ as the query
formed by joining the relations in $\lambda(v)$. The {\bf {\em
    (fractional) width}} of a GHD $D$ is $\mathsf{AGM}(Q_v)$, which is
an upper bound on the number of tuples returned by $Q_v$. The {\bf
  {\em fractional hypertree width (fhw)}} of a hypergraph $H$ is the
minimum width of all GHDs of $H$. Given a GHD with width $w$, there is
a simple algorithm to run in time $O(N^{w} + {\out})$. First, run any
worst-case optimal algorithm on $Q_v$ for each node $v$ of the GHD;
each join takes time $O(N^{w})$ and produces at most $O(N^{w})$
tuples. Then, one is left with an acyclic query over the output of
$Q_v$, namely the tree itself. We then perform Yannakakis' classical
algorithm~\cite{DBLP:conf/vldb/Yannakakis81}, which for acyclic
queries enables us to compute the output in linear time in the input
size ($O(N^w)$) plus the output size ($\out$).

\subsection{Choosing Logical Query Plans}
\label{sec:query_plans}

We describe how \EH chooses GHDs, explain how we leverage previous work to enable aggregations over GHDs, and describe how GHDs are used to select a global attribute ordering in \EH{}. In \Cref{sec:selections}, we provide detail on how classic database optimizations, such as pushing down selections, can be captured using GHDs.

\paragraph*{GHD Optimizer}
The \EH query compiler selects an optimal GHD to ensure tighter theoretical run time guarantees. 
It is key that the \EH optimizer selects a GHD with the
smallest width $w$ to ensure an optimal GHD. Similar to how a traditional database
pushes down projections to minimize the output size, \EH minimizes the
output size by finding the GHD with the smallest width. In contrast to pushing
down projections, finding the minimum width GHD is $\mathsf{NP}$-hard
in the number of relations and attributes. As the number of relations and attributes is typically small (three for triangle counting), we simply brute force search GHDs of all possible widths.

\paragraph*{Aggregations over GHDs} Previous work has investigated aggregations over hypertree decompositions \cite{olteanu2015size,gottlob2005hypertree}. \EH adopts this previous work in a straightforward way. To do this, we add a single attribute with ``semiring annotations'' following Green et al.~\cite{green2007provenance}. \EH simply manipulates this value as it is projected away. This general notion of aggregations over annotations enables \EH to support traditional notions of queries with aggregations as well as a wide range of workloads outside traditional data processing, like message passing in graphical models.

\paragraph*{Global Attribute Ordering}
Once a GHD is selected, \EH selects a global attribute ordering. The global attribute ordering determines the order in which \EH code generates the generic worst-case optimal algorithm (\Cref{fig:worst_case}) and the index structure of our tries (\Cref{sec:input_data}). Therefore, selecting a global attribute ordering is analogous to selecting a join and index order in a traditional pairwise relational engine. The attribute order depends on the query. For the purposes of this paper, we assume both trie orderings are present, and we are therefore free to select any attribute order. For graphs (two-attributes), most in-memory graph engines maintain both the matrix and its transpose in the compressed sparse row format \cite{Hong:2012:GDE:2150976.2151013,Nguyen:2013:LIG:2517349.2522739}. We are the first to consider selecting an attribute ordering based on a GHD and as a result we explore simple heuristics based on structural properties of the GHD. To assign an attribute order for all queries in this paper, \EH simply performs a pre-order traversal over the GHD, adding the attributes from each visited GHD node into a queue. 

\subsection{Code Generation}
\label{sec:qp_codegen}
\EH's code generator converts the selected GHD for each query into optimized C++ code that uses the operators in \Cref{lst:eh_operators}. We choose to implement code generation in \EH as it is has been shown to be an efficient technique to translate high-level query plans into code optimized for modern hardware \cite{neumann2011efficiently}. 

\subsubsection{Code Generation API}
We first describe the storage-engine operations which serve as the basic high-level API for our generated code. Our trie data structure offers a standard, simple API for traversals and set intersections that is sufficient to express the worst-case optimal join algorithm detailed in Algorithm~\ref{fig:worst_case}.  The key operation over the trie is to return a set of values that match a specified tuple predicate (see \Cref{lst:eh_operators}). This operation is typically performed  while traversing the trie, so \EH{} provides an optimized iterator interface. The set of values retrieved from the trie can be intersected with other sets or iterated over using the operations in \Cref{lst:eh_operators}. 

\begin{table}
  \small
  \centering
  \begin{tabular}{lll}
  \toprule
   & Operation & Description \\
  \midrule
  \multirow{4}{*}{Trie ($R$)} & \multirow{2}{*}{$R$[$t$]} & Returns the set \\
  & & matching tuple $t \in R$. \\
  & \multirow{2}{*}{$R \leftarrow R \cup t \times xs$} & Appends elements in set $xs$ \\
  & &  to tuple $t \in R$. \\
  \midrule
  \multirow{4}{*}{Set ($xs$)} & \multirow{2}{*}{for $x$ in $xs$} & Iterates through the \\
  & & elements $x$ of a set $xs$. \\
  & \multirow{2}{*}{$xs$ $\cap$ $ys$} & Returns the intersection \\
  & & of sets $xs$ and $ys$. \\
  \bottomrule
  \end{tabular}
  \caption{Execution Engine Operations}
  \label{lst:eh_operators}
\end{table}

\vspace{5mm}
\subsubsection{GHD Translation}
The goal of code generation is to translate a GHD to the operations in \Cref{lst:eh_operators}. Each GHD node $v \in V(T)$
is associated with a trie described by the attribute ordering in
$\chi(v)$. Unlike previous worst-case optimal join engines, there are two phases to our algorithm: (1) within nodes of
$V(T)$ and (2) between nodes $V(T)$.

\paragraph*{Within a Node} For each $v \in V(T)$, we run the generic worst-case optimal algorithm shown in
\Cref{fig:worst_case}. Suppose $Q_v$ is the triangle query.
 \begin{example}
   Consider the triangle query. The hypergraph is $V= \{ X,Y,Z\}$ and $E = \{R,S,T\}$. In the
first call, the loop body generates a loop
with body $\textrm{Generic-Join}(\\ \{Y,Z\}, E, t_{X})$. In turn, with two more
calls this generates:

\begin{tabbing}
for \= $t_{X} \in \pi_{X}R \cap \pi_{X}T$ do\\
\>for \=$t_{Y} \in \pi_{Y} R[t_X] \cap \pi_{Y} S$ do\\
\>\>$Q \leftarrow Q \cup (t_x,t_y) \times \left(\pi_{Z} S[t_Y] \cap \pi_{Z} T[t_X]\right)$.
\end{tabbing}
\end{example}

 \paragraph*{Across Nodes}
 Recall Yannakakis' seminal
 algorithm~\cite{DBLP:conf/vldb/Yannakakis81}: we first perform a
 ``bottom-up'' pass, which is a reverse level-order traversal of
 $T$. For each $v \in V(T)$, the algorithm computes $Q_v$ and passes its results to the parent node. Between nodes $(v_0,v_1)$ we
 pass the relations projected onto the shared attributes $\chi(v_0)
 \cap \chi(v_1)$. Then, the result is constructed by
 walking the tree ``top-down'' and collecting each result.

\paragraph*{Recursion} \EH supports both naive and seminaive
  evaluation to handle recursion. For naive recursion, \EH's optimizer produces a
  (potentially infinite) linear chain GHD with the output of one GHD
  node serving as the input to its parent GHD node. We run naive
  recursion for PageRank in \Cref{lst:eh_query_language}. This boils to down to a simple unrolling of the join algorithm. Naive recursion is not an acceptable solution in applications such as SSSP where work is continually being eliminated. To
  detect when \EH should run seminaive recursion, we check if the
  aggregation is monotonically increasing or decreasing with a \texttt{MIN} or \texttt{MAX}
  operator. We use seminaive recursion for SSSP.

 \begin{example} For the
 Barbell query (see \Cref{fig:barbell_ghd}), we first run \Cref{fig:worst_case} on nodes $v_1$ and $v_2$; 
 then we project their results on $x$ and $x'$ and pass them to node
$v_0$. This is part of the ``bottom-up'' pass. We then execute \Cref{fig:worst_case} on node $v_0$ which now contains the results (triangles) of its children. \Cref{fig:worst_case} executes here by simply checking for pairs of $(x,x')$ from its children that are in $U$. To perform the ``top-down'' pass, for each matching pair, we append $(y,z)$ from $v_1$ and $(y',z')$ from $v_2$.
 \end{example}

\section{Execution Engine Optimizer}
\label{sec:execution_engine}

The \EH execution engine runs code generated from the query compiler. The goal of the \EH execution engine is to fully utilize SIMD parallelism, but extracting SIMD parallelism is challenging as graph data is often skewed in several distinct ways. The density of data values is almost never constant: some parts of the relation are dense while others are sparse.  We call this \emph{density skew}.\footnote{We measure density skew using the Pearson's first coefficient of skew defined as $3\sigma^{-1}(mean-mode)$ where $\sigma$ is the standard deviation.} A novel aspect of \EH is that it automatically copes with density skew through an optimizer that selects among different data layouts. We implemented and tested five different set layouts previously proposed in the literature \cite{inoue2014faster,highlyscalable,Schlegel11fastsorted-set,DBLP:journalscorrLemireBK14}. We found that the simple \uint and \bitset layouts yield the highest performance in our experiments \cite{DBLP:journals/corr/AbergerNOR15}. Thus, we focus on selecting between (1) a 32-bit unsigned integer (\uint) layout for sparse data and (2) a \bitset layout for dense data.  For dense data, the \bitset layout makes it trivial to take advantage of SIMD parallelism. But for sparse data, the \bitset layout causes a quadratic blowup in memory usage while \uint sets make extracting SIMD parallelism challenging. 

Making these layout choices is challenging, as the optimal choice depends both on the characteristics of the data, such as density, and the characteristics of the query. We first describe layouts and intersection algorithms in \Cref{sec:layouts,sec:intersections}. This serves as background for the tradeoff study we perform in \Cref{sec:graph_level}, where we explore the proper granularity at which to make layout decisions. Finally, we present our automatic optimizer and show that it is close to an unachievable lower-bound optimal in \Cref{sec:optimizer_gran}. This study serves as the basis for our automatic layout optimizer that we use inside of the \EH storage engine.

\begin{figure}
  \small
  \centering
  \begin{tabular}{@{}|l|l|l|l|l|l|l|l|l|l|@{}}
  \hline
  $n$ & $o_1$ & \dots & $o_n$ & $b_1$ & \dots & $b_n$ \\
  \hline
  \end{tabular}
  \caption{Example of the \bitset layout that contains $n$
  blocks and a sequence of offsets ($o_1$-$o_n$) and blocks ($b_1$-$b_n$). The offsets store the start offset for values in the bitvector.}
\label{fig:bs}
\end{figure}

\subsection{Layouts}
\label{sec:layouts}
In the following, we describe the \bitset layout in \EH. We omit a description of the \uint layout as it is just an array of 32-bit unsigned integers. We also detail how both layouts support associated data values.  

\paragraph*{BITSET}
The \bitset layout stores a set of pairs (offset, bitvector), as
shown in \Cref{fig:bs}. The offset is the index of the smallest
value in the bitvector. Thus, the layout is a compromise
between sparse and dense layouts. We refer to the number of
bits in the bitvector as the {\em block size}. \EH{}
supports block sizes that are powers of two with a default of 256.\footnote{
The width of an AVX register.} 
As shown, we pack the offsets contiguously, which allows us to
regard the offsets as a \uint layout; in turn, this allows \EH
to use the same algorithm to intersect the offsets as it does for the
\uint layout. 

\paragraph*{Associated Values}
Our sets need to be able to store associated values such as pointers
to the next level of the trie or annotations of arbitrary types. In \EH, the associated values for each set also use different underlying data layouts based on the type of the underlying set. For the \bitset layout we store the associated values as a dense vector (where associated values are accessed based upon the data value in the set). For the \uint layout we store the associated values as a sparse vector (where the associated values are accessed based upon the index of the value in the set).

\begin{figure}
\hfill
\parbox[t]{.23\textwidth}{
  \centering
  \includegraphics[width=\linewidth]{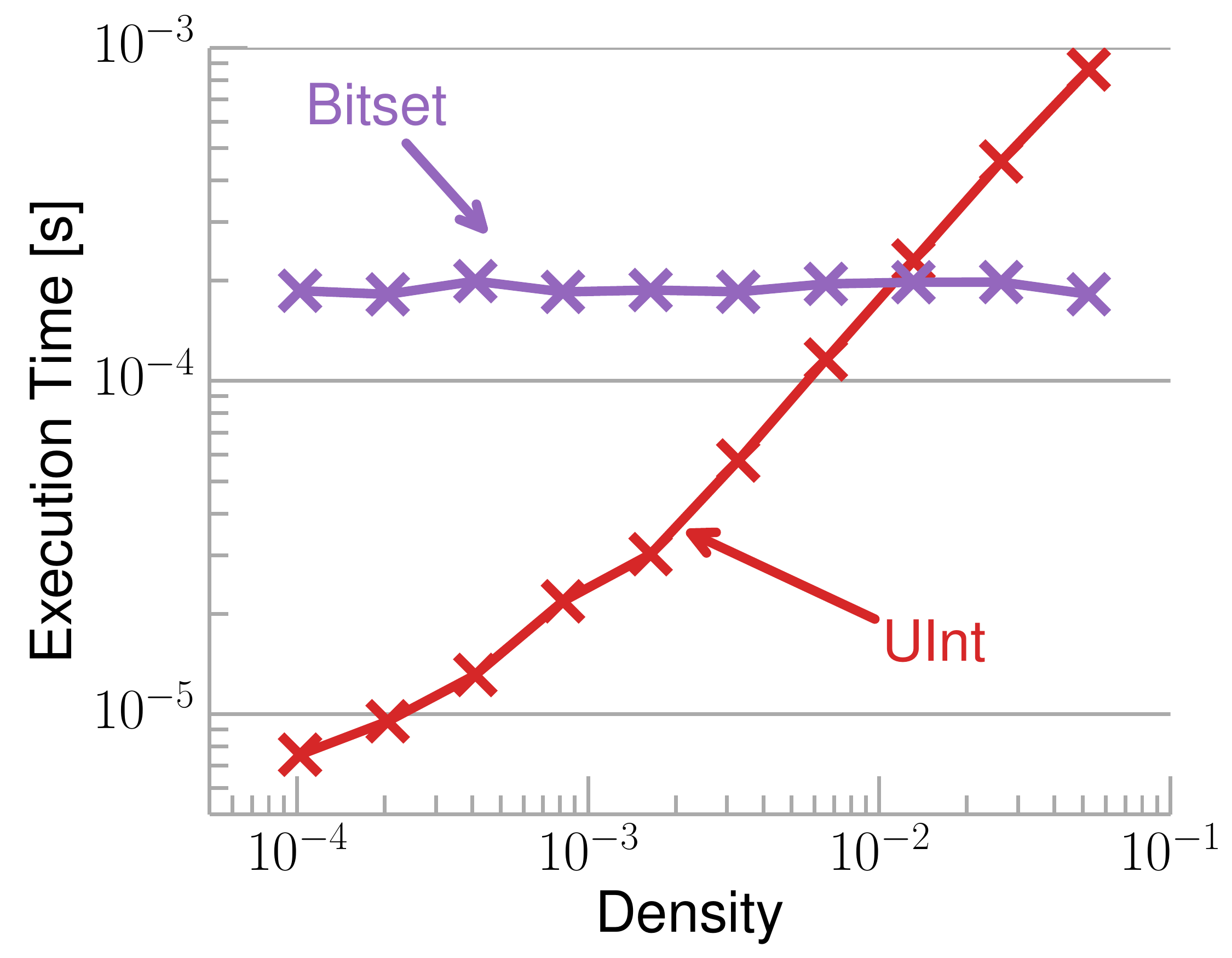}
  \caption{Intersection time of \uint and \bitset layouts for different densities.}
  \label{fig:impact_density}
}
\hfill
\parbox[t]{.23\textwidth}{
  \centering
  \includegraphics[width=\linewidth]{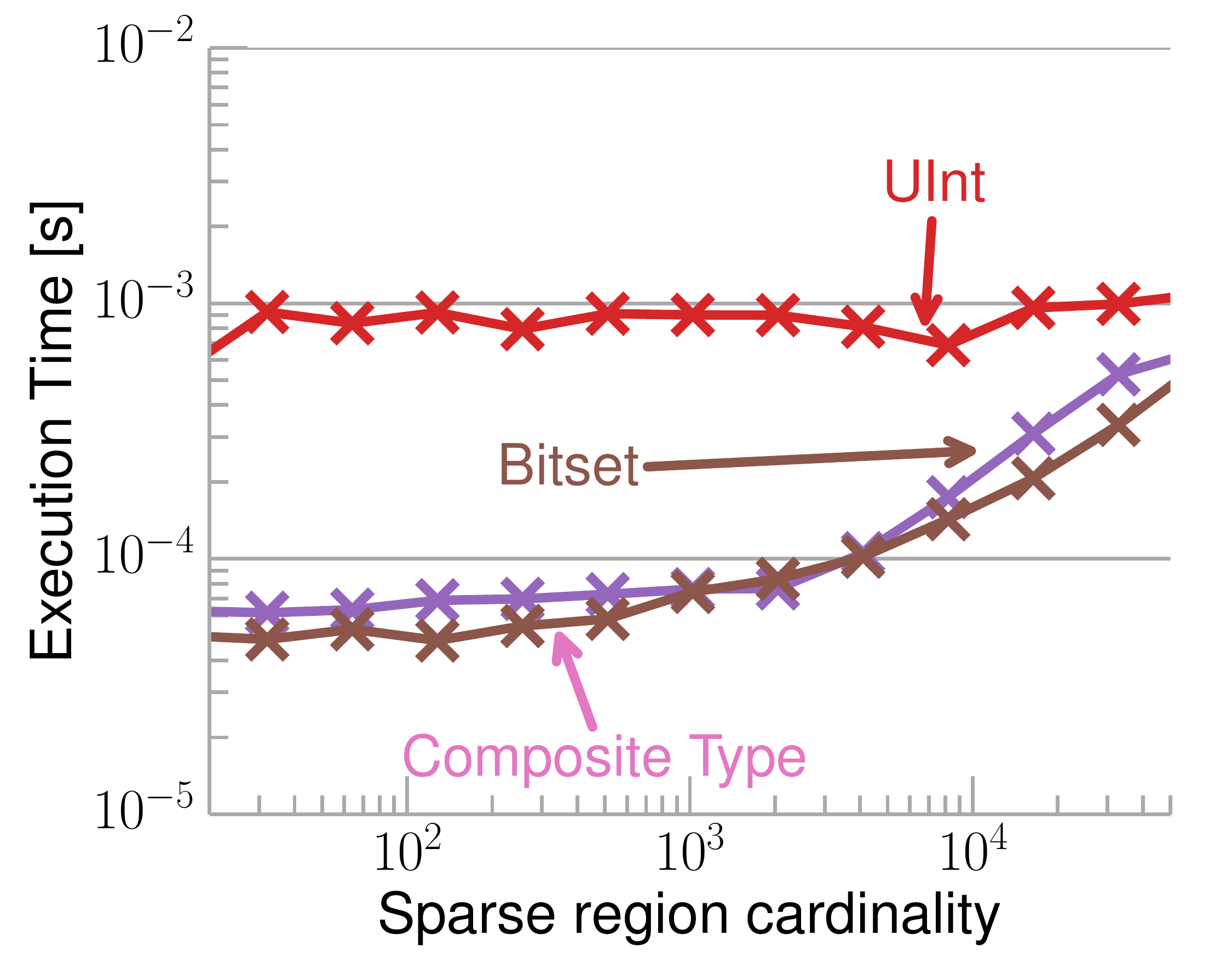}
  \caption{Intersection time of layouts for sets with different sizes of dense regions.}
  \label{fig:skew}
}
\end{figure}

\subsection{Intersections}
\label{sec:intersections}

We briefly present an overview of the intersection algorithms \EH uses for each layout. This serves as the background for our tradeoff study in \Cref{sec:graph_level}. We remind the reader that the  {\em min property} presented in \Cref{sec:wc_back} must hold for set intersections 
so that a worst-case optimal runtime can be guaranteed in \EH.

\paragraph*{UINT $\cap$ UINT} For the \uint layout, we implemented and tested five state-of-the-art SIMD set intersections \cite{DBLP:journalscorrLemireBK14,inoue2014faster,Schlegel11fastsorted-set,highlyscalable}. For \uint intersections we found that the size of two sets being intersected may be drastically different. This is another type of skew, which we call \emph{cardinality skew}. So-called {\em galloping} algorithms~\cite{veldhuizen2012leapfrog} allow one to run in time proportional to the size of the smaller set, which copes with cardinality skew. However, for sets that are of similar size, galloping algorithms may have additional overhead. Therefore, like others \cite{DBLP:journalscorrLemireBK14,inoue2014faster}, \EH uses a simple hybrid algorithm that selects a SIMD galloping algorithm when the ratio of cardinalities is greater than 32:1, and a SIMD shuffling algorithm otherwise. 

\paragraph*{BITSET $\cap$ BITSET} Our \bitset is conceptually a two-layer structure of offsets and blocks. Offsets are stored using \uint sets. Each offset determines the start of the corresponding block. To compute the intersection, we first find the common blocks between the \bitsets by intersecting the offsets using a \uint intersection followed by SIMD {\tt AND} instructions to intersect matching blocks. In the best case, i.e., when all bits in the register are $1$, a single hardware instruction computes the intersection of 256 values.

\paragraph*{UINT $\cap$ BITSET}
To compute the intersection between a \uint and a \bitset, we first intersect the \uint values with the offsets in the \bitset. We do this to check if it is possible that some value in a \bitset block matches a \uint value. As \bitset block sizes are powers of two in \EH{}, this can be accomplished by masking out the lower bits of each \uint value in the comparison. This check may result in false positives, so, for each matching \uint and \bitset block we check whether the corresponding \bitset blocks contain the \uint value by probing the block. We store the result as a \uint as the intersection of two sets can be at most as dense as the sparser set.\footnote{
Estimating data characteristics like output
cardinality a priori is a hard problem~\cite{chaudhuri1999random}
and we found it is too costly to reinspect the data after each operation.} Notice that this algorithm satisfies the min property with a constant determined by the block size.

\vspace{14mm}
\subsection{Tradeoffs}
\label{sec:graph_level}

We explore three different levels of granularity to decide between \uint and \bitset layouts in our trie data structure:
the relation level, the set level, and the block level. 

\paragraph*{Relation Level} Set layout decisions at the relation level force the data in all relations to be stored using the same layout and therefore do not address
density skew. The simplest layout in memory is to store
all sets in every trie using the \uint layout. Unfortunately, it is
difficult to fully exploit SIMD parallelism using this
layout, as only four elements fit in a single SIMD
register.\footnote{In the Intel Ivy Bridge architecture only SSE instructions contain integer comparison mechanisms;
therefore we are forced to restrict ourselves to a 128 bit register width.} In
contrast, bitvectors can store 256 elements in a single SIMD register. However,
bitvectors are inefficient on sparse data and can result in a quadratic blowup
of memory usage. Therefore, one would expect unsigned integer arrays to be well
suited for sparse sets and bitvectors for dense sets. \Cref{fig:impact_density} illustrates
this trend. Because of the sparsity in real-world data, we found that \uint provides the best performance at the relation level.

\paragraph*{Set Level} Real-world data often has a large amount of density skew, so both the \uint and
\bitset layouts are useful.
At the set level we simply decide on a per-set level if the entire set should be represented using a \uint or a \bitset layout. Furthermore, we found that our \uint and \bitset intersection
can provide up to a 6x performance increase over the best homogeneous \uint
intersection and a 132x increase over a homogeneous \bitset intersection.  We
show in \Cref{sec:eh_optimizations,sec:optimizer_gran} that the impact of mixing
layouts at the set level on real data can increase overall query performance by over an order of magnitude.

\paragraph*{Block Level} Selecting a layout at the set level might be too coarse if there is
internal skew. For example, set level layout decisions are too coarse-grained to optimally exploit a set with a large sparse region followed by a
dense region. Ideally, we would like to treat dense regions separately from
sparse ones. To deal with skew at a finer granularity, we propose a {\em
composite set} layout that regards the domain as a series of
fixed-sized blocks; we represent sparse blocks using the \uint layout
and dense blocks using the \bitset layout. We show in \Cref{fig:skew} that on synthetic data the composite layout can outperform the \uint and \bitset layouts by 2x. 


\begin{table}
  \small
  \begin{center}
    \begin{tabular}{@{}lrrrrp{2.0cm}@{}}
    \toprule
    Dataset & \parbox[c]{0.7cm}{Nodes [M]} & \parbox[c]{1.0cm}{Dir. Edges [M]}  & \parbox[c]{0.7cm}{Undir. Edges [M]} & \parbox[c]{0.7cm}{Density Skew} & Description \\
    \midrule
    Google+\cite{snapnets} & 0.11 & 13.7 & 12.2 & 1.17 & User network \\
    Higgs\cite{snapnets} & 0.4 & 14.9 & 12.5 & 0.23 & Tweets about Higgs Boson \\
    LiveJournal\cite{konect:leskovec08} & 4.8 & 68.5 & 43.4 & 0.09 & User network \\
    Orkut\cite{konect:mislove} & 3.1 & 117.2 & 117.2 & 0.08 & User network \\
    Patents\cite{konect:2014:patentcite} & 3.8 & 16.5 & 16.5 & 0.09 & Citation \blank{0.4cm} network \\
    Twitter\cite{konect:twitter1} & 41.7 & 1,468.4 & 757.8 & 0.12 & Follower \blank{0.4cm} network \\
    \bottomrule

    \end{tabular}
    \caption{Graph datasets presented in \Cref{sec:workloads} that are used in the experiments.}
    \label{table:datasets}
  \end{center}
\end{table}

\subsection{Layout Optimizer}
\label{sec:optimizer_gran}

Our synthetic experiments in \Cref{sec:graph_level} show there is no clear winner, as the right granularity at which to make a layout decision depends on the data characteristics and the query. To determine if our system should make layout decisions at a relation, set, or block level on real data, we compare each approach to the time of a lower-bound oracle optimizer. We found that while running on the real graph datasets shown in \Cref{table:datasets}, choosing layouts at the set level provided the best overall performance (see \Cref{table:oracle}).

\begin{table}
  \small
  \begin{center}
  \begin{tabular}{@{}lrrr@{}}
    \hline
    \bigstrut
    Dataset      & Relation level & Set level & Block level \\
    \hline
    \bigstrut[h]
    Google+      & 7.3x & 1.1x & 3.2x \\
    Higgs        & 1.6x & 1.4x & 2.4x \\
    LiveJournal  & 1.3x & 1.4x & 2.0x \\
    Orkut        & 1.4x & 1.4x & 2.0x \\
    Patents      & 1.2x & 1.6x & 1.9x \\
    \hline
 \end{tabular}
 \end{center}
 \caption{Relative time of the level optimizers on triangle counting compared to the oracle.}
 \label{table:oracle}
\end{table}

\paragraph*{Oracle Comparison}  The oracle optimizer we compare to provides a lower bound as it is able to freely select amongst all layouts per set operation. Thus, it is allowed to choose any layout and intersection combination while assuming perfect knowledge of the cost of each intersection. We implement the oracle optimizer by brute-force, running all possible layout and algorithm combinations for every set intersection in a given query. The oracle optimizer then counts only the cost of the best-performing combination (from all possible combinations), therefore providing a lower bound for the \EH optimizer. On the triangle counting query, the set level optimizer was at most 1.6x off the optimal oracle performance, while choosing at the relation and block levels can be up to 7.3x and 3.2x slower respectively than the oracle. Although more sophisticated optimizers exist, and were tested in the \EH engine, we found that this simple set level optimizer performed within 10\%-40\% of the oracle optimizer on real graph data. Because of this we use the set optimizer by default inside of \EH (and for the remainder of this paper).

\paragraph*{Set Optimizer} The set optimizer in \EH selects the layout for a set in isolation based on its cardinality and range. It selects the \bitset layout when each value in the set consumes at most as much space as a SIMD (AVX) register and the \uint layout otherwise. The optimizer uses the \bitset layout with a block size equal to the range of the data in the set. We find this to be more effective than a fixed block size since it lacks the overhead of storing multiple offsets.

\section{Experiments}
\label{sec:experiments}
We compare \EH against state-of-the-art high- and low-level specialized graph engines on standard graph benchmarks. We show that by using our optimizations from \Cref{sec:qc}  and \Cref{sec:execution_engine}, \EH is able to compete with specialized graph engines.

\subsection{Experiment Setup}

We describe the datasets, comparison engines, metrics, and experiment setting used to validate that \EH competes with specialized engines in \Cref{sec:experiment_comparison,sec:eh_optimizations}. 

\subsubsection{Datasets}
\label{sec:workloads}

\Cref{table:datasets} provides a list of the 6 popular datasets that we use in our comparison to other graph analytics engines. LiveJournal, Orkut, and Patents are graphs with a low amount of density skew, and Patents is much smaller graph in comparison to the others. Twitter is one of the largest publicly available datasets and is a standard benchmarking dataset that contains a modest amount of density skew. Higgs is a medium-sized graph with a modest amount of density skew. Google+ is a graph with a large amount of density skew.

\subsubsection{Comparison Engines}
We compare \EH against popular high- and low-level engines in the graph domain. We also compare to the high-level LogicBlox engine, as it is the first commercial database with a worst-case optimal join optimizer. 

\paragraph*{Low-Level Engines} We benchmark several graph analytic engines and compare their performance. The engines that we compare to are PowerGraph v2.2 \cite{Gonzalez:2012:PDG:2387880.2387883}, the latest release of commercial graph tool (CGT-X), and Snap-R \cite{snap}.  Each system provides highly optimized shared memory implementations of the triangle counting query.  Other shared memory graph engines such as Ligra \cite{Shun:2013:LLG:2517327.2442530} and Galois \cite{Nguyen:2013:LIG:2517349.2522739} do not provide optimized implementations of the triangle query
and requires one to write queries by hand. We do provide a comparison to Galois v2.2.1 on PageRank and SSSP. Galois has been shown to achieve performance similar to that of Intel's hand-coded implementations \cite{Satish:2014:NMG:2588555.2610518} on these queries.  

\paragraph*{High-Level Engines} We compare to LogicBlox v4.3.4 on all queries since LogicBlox is the first general purpose commercial engine to provide similar worst-case optimal join guarantees. LogicBlox also provides a relational model that makes complex queries easy and succinct to express. It is important to note that LogicBlox is full-featured commercial system (supports transactions, updates, etc.) and therefore incurs inefficiencies that \EH{} does not. Regardless, we demonstrate that using GHDs as the intermediate representation in \EH's query compiler not only provides tighter theoretical guarantees, but provides more than a three orders of magnitude performance improvement over LogicBlox. We further demonstrate that our set layouts account for over an order of magnitude performance advantage over the LogicBlox design. We also compare to SociaLite \cite{seo2013socialite} on each query as it also provides high-level language optimizers, making the queries as succinct and easy to express as in \EH. Unlike LogicBlox, SociaLite does not use a worst-case optimal join optimizer and therefore suffers large performance gaps on graph pattern queries. Our experimental setup of the LogicBlox and SociaLite engines was verified by an engineer from each system and our results are in-line with previous findings \cite{seo2013socialite,nguyen2015join,Satish:2014:NMG:2588555.2610518}.

\paragraph*{Omitted Comparisons} We compared \EH to \\GraphX \cite{Xin:2013:GRD:2484425.2484427} which is a graph engine designed for scale-out performance. GraphX was consistently several orders of magnitude slower than \EH's performance in a shared-memory setting. We also compared to a commercial database and PostgreSQL but they were consistently over three orders of magnitude off of \EH's performance. We exclude a comparison to the Grail method \cite{fancase} as this approach in a SQL Server has been shown to be comparable to or sometimes worse than PowerGraph \cite{Gonzalez:2012:PDG:2387880.2387883} when the entire dataset can easily fit in-memory (like we consider in this paper). It should be noted that the Grail method with a persistent database has been shown to be more robust than in-memory engines, such as \EH and PowerGraph, when the entire dataset does not fit easily in-memory \cite{fancase}.

\subsubsection{Metrics}
 We measure the performance of \EH{} and other engines. For end-to-end performance, we measure the wall-clock time for each system to complete each query. This measurement excludes the time used for data loading, outputting the result, data statistics collection, and index creation for all engines. We repeat each measurement seven times, eliminate the lowest and the highest value, and report the average. Between each measurement of the \emph{low-level} engines we wipe the caches and re-load the data to avoid intermediate results that each engine might store. For the \emph{high-level} engines we perform runs back-to-back, eliminating the first run which can be an order of magnitude worse than the remaining runs. We do not include compilation times in our measurements. Low-level graph engines run as a stand-alone program (no compilation time) and we discard the compilation time for high-level engines (by excluding their first run, which includes compilation time). Nevertheless, our unoptimized compilation process (under two seconds for all queries in this paper) is often faster than other high-level engines' (Socialite or LogicBlox).

\subsubsection{Experiment Setting}
\label{sec:experiment_setting}

\EH is an in-memory engine that runs and is evaluated on a single node server.
As such, we ran all experiments on a single machine with a total of 48 cores on four Intel Xeon E5-4657L v2 CPUs and 1 TB of RAM. We compiled the C++ engines (\EH{}, Snap-R, PowerGraph, TripleBit) with g++ 4.9.3 (-O3) and ran the Java-based engines (CGT-X, LogicBlox, SociaLite) on OpenJDK 7u65 on Ubuntu 12.04 LTS. For all engines, we chose buffer and heap sizes that were at least an order of magnitude larger than the dataset itself to avoid garbage collection.

\subsection{Experimental Results}
\label{sec:experiment_comparison}
We provide a comparison to specialized graph analytics engines on several standard workloads. We demonstrate that \EH outperforms the graph analytics engines by 2-60x on graph pattern queries while remaining competitive on PageRank and SSSP.

\subsubsection{Graph Pattern Queries}
\label{sec:triangle}

\begin{table}
  \setlength{\tabcolsep}{3pt}
  \small
  \begin{center}
    \begin{tabular}{@{}lrrrrrr@{}}
    \toprule
    & &  \multicolumn{3}{c}{Low-Level} & \multicolumn{2}{c}{High-Level}  \\
    \cmidrule(r){3-5}     \cmidrule(r){6-7}
    Dataset & \multicolumn{1}{c}{\EHSMALL} & \multicolumn{1}{c}{PG} & \multicolumn{1}{c}{CGT-X} & \multicolumn{1}{c}{SR} & \multicolumn{1}{c}{SL} & \multicolumn{1}{c}{LB} \\
    \midrule          
    Google+           &\textbf{0.31}      &8.40x        &62.19x        &4.18x	&1390.75x    &83.74x\\    
    Higgs             &\textbf{0.15}      &3.25x       &57.96x        &5.84x	&387.41x     &29.13x\\    
    LiveJournal       &\textbf{0.48}      &5.17x       &3.85x         &10.72x	&225.97x    &23.53x\\    
    Orkut             &\textbf{2.36}      &2.94x       &-             &4.09x 	&191.84x    &19.24x\\   
    Patents           &\textbf{0.14}      &10.20x      &7.45x         &22.14x	&49.12x    &27.82x\\    
    Twitter           &\textbf{56.81}     &4.40x       &-             &2.22x  	&t/o   &30.60x\\      
    \bottomrule

    \end{tabular}

    \caption{ Triangle counting runtime (in seconds) for \EH (\EHSMALL) and relative slowdown for other engines including PowerGraph (PG), a commercial graph tool (CGT-X), Snap-Ringo (SR), SociaLite (SL) and LogicBlox (LB). 48 threads used for all engines. ``-'' indicates the engine does not process over 70 million edges. ``t/o'' indicates the engine ran for over 30 minutes.} 

    \label{table:tcount}
  \end{center}
\end{table}

We first focus on the triangle counting query as it is a standard graph pattern benchmark with hand-tuned implementations provided in both high- and low-level engines. Furthermore, the triangle counting query is widely used in graph processing applications and is a common subgraph query pattern \cite{newman2003structure,milo2002network}. To be fair to the low-level frameworks, we compare the triangle query only to frameworks that provide a hand-tuned implementation. Although we have a high-level optimizer, we outperform the graph analytics engines by 2-60x on the triangle counting query.

As is the standard, we run each engine on pruned
versions of these datasets, where each undirected edge is pruned such that $src_{id} > dst_{id}$ and $id$'s are assigned based upon the degree of the node. This process is standard as it limits the size of the intersected sets and has been shown to empirically work well \cite{Schank:2005:FCL:2154763.2154820}. Nearly every graph engine implements pruning in this fashion for the triangle query.

\paragraph*{Takeaways}

The results from this experiment are in \Cref{table:tcount}. On very sparse datasets with low density skew (such as the Patents dataset) our performance gains are modest as it is best to represent all sets in the graph using the \uint layout, which is what many competitor engines  already do. As expected, on datasets with a larger degree of density skew, our performance gains become much more pronounced. For example, on the Google+ dataset, with a high density skew, our set level optimizer selects 41\% of the neighborhood sets to be \bitsets and achieves over an order of magnitude performance gain over representing all sets as \uints. LogicBlox performs well in comparison to CGT-X on the Higgs dataset, which has a large amount of cardinality skew, as they use a Leapfrog Triejoin algorithm \cite{veldhuizen2012leapfrog} that optimizes for cardinality skew by obeying the min property of set intersection. \EH similarly obeys the min property by selecting amongst set intersection algorithms based on cardinality skew. In \Cref{sec:eh_optimizations} we demonstrate that over a two orders of magnitude performance gain comes from our set layout and intersection algorithm choices. 

\paragraph*{Omitted Comparison}
 We do not compare to Galois on the triangle counting query, as Galois does not provide an implementation and implementing it ourselves would require us to write a custom set intersection in Galois (where >95\% of the runtime goes). We describe how to implement high-performance set intersections in-depth in \Cref{sec:execution_engine} and \EH's triangle counting numbers are comparable to Intel's hand-coded numbers which are slightly (10-20\%) faster than the Galois implementation \cite{Satish:2014:NMG:2588555.2610518}. We provide a comparison to Galois on SSSP and PageRank in \Cref{sec:graph_analytics}. 

\subsubsection{Graph Analytics Queries}
\label{sec:graph_analytics}
Although \EH is capable of expressing a variety of different workloads, we benchmark PageRank and SSSP as they are common graph benchmarks. In addition, these benchmarks illustrate the capability of \EH to process broader workloads that relational engines typically do not process efficiently: (1) linear algebra operations (in PageRank) and (2) transitive closure (in SSSP). We run each query on undirected versions of the graph datasets and demonstrate competitive performance compared to specialized graph engines. Our results suggest that our approach is competitive outside of classic join workloads.

\paragraph*{PageRank}  As shown in \Cref{table:pr}, we are consistently 2-4x faster than standard low-level baselines and more than an order of magnitude faster than the high-level baselines on the PageRank query. We observe competitive performance with Galois (271 lines of code), a highly tuned shared memory graph engine, as seen in \Cref{table:pr}, while expressing the query in three lines of code (\Cref{lst:eh_query_language}). There is room for improvement on this query in \EH since double buffering and the elimination of redundant joins would enable \EH to achieve performance closer to the bare metal performance, which is necessary to outperform Galois. 

\paragraph*{Single-Source Shortest Paths} We compare \EH's performance to LogicBlox and specialized engines in \Cref{table:sssp} for SSSP while omitting a comparison to Snap-R. Snap-R does not implement a parallel version of the algorithm and is over three orders of magnitude slower than \EH on this query. For our comparison we selected the highest degree node in the undirected version of the graph as the start node. \EH consistently outperforms PowerGraph (low-level) and SociaLite (high-level) by an order of magnitude and LogicBlox by three orders of magnitude on this query. More sophisticated implementations of SSSP than what \EH generates exist \cite{Beamer:2012:DBS:2388996.2389013}. For example, Galois, which implements such an algorithm, observes a 2-30x performance improvement over \EH on this application (\Cref{table:sssp}). Still, \EH is competitive with Galois (172 lines of code) compared to the other approaches while expressing the query in two lines of code (\Cref{lst:eh_query_language}).

\begin{table}
  \small
  \begin{center}
    \setlength{\tabcolsep}{3pt}
    \begin{tabular}{@{}lrrrrrrr@{}}
    \toprule
    & &  \multicolumn{4}{c}{Low-Level} & \multicolumn{2}{c}{High-Level}  \\
    \cmidrule(r){3-6}     \cmidrule(r){7-8}
    Dataset & \multicolumn{1}{c}{\EHSMALL} & \multicolumn{1}{c}{G} & \multicolumn{1}{c}{PG} & \multicolumn{1}{c}{CGT-X} & \multicolumn{1}{c}{SR} & \multicolumn{1}{c}{SL} & \multicolumn{1}{c}{LB} \\
    \midrule          
    Google+           &0.10    &\textbf{0.021}     &0.24       &1.65         &0.24	&1.25	&7.03    \\    %
    Higgs             &0.08    &\textbf{0.049}     &0.5        &2.24         &0.32	&1.78	&7.72    \\    %
    LiveJournal       &0.58    &\textbf{0.51}      &4.32       &-            &1.37	&5.09	&25.03   \\    %
    Orkut             &0.65    &\textbf{0.59}      &4.48       &-            &1.15	&17.52	&75.11   \\    %
    Patents           &\textbf{0.41}    &0.78      &3.12       &4.45         &1.06	&10.42	&17.86   \\    %
    Twitter           &\textbf{15.41}   &17.98     &57.00      &-            &27.92	&367.32	&  442.85   \\    %
    \bottomrule

    \end{tabular}

    \caption{ Runtime for 5 iterations of PageRank (in seconds) using 48 threads for all engines. ``-'' indicates the engine does not process over 70 million edges. \EHSMALL denotes \EH and the other engines include Galois (G), PowerGraph (PG), a commercial graph tool (CGT-X), Snap-Ringo (SR), SociaLite (SL), and LogicBlox (LB).}

    \label{table:pr}
  \end{center}
\end{table}

\begin{table}
  \small
    \setlength{\tabcolsep}{3pt}
  \begin{center}
    \begin{tabular}{@{}lrrrrrr@{}}
    \toprule
    & &  \multicolumn{3}{c}{Low-Level} & \multicolumn{2}{c}{High-Level}  \\
    \cmidrule(r){3-5} \cmidrule(r){6-7}
    Dataset & \multicolumn{1}{c}{\EHSMALL} &\multicolumn{1}{c}{G} & \multicolumn{1}{c}{PG} & \multicolumn{1}{c}{CGT-X} & \multicolumn{1}{c}{SL} & \multicolumn{1}{c}{LB} \\
    \midrule          
    Google+           &0.024    &\textbf{0.008}    &0.22    	&0.51	&0.27  	&41.81	\\  
    Higgs             &0.035    &\textbf{0.017}    &0.34    	&0.91	&0.85 	&58.68	\\  
    LiveJournal       &0.19    	&\textbf{0.062}    &1.80    	&- 		&3.40	&102.83 	\\   
    Orkut             &0.24    	&\textbf{0.079}    &2.30    	&- 		&7.33 	&215.25	\\   
    Patents           &0.15    	&\textbf{0.054}    &1.40    	&4.70 	&3.97 	&159.12	\\   
    Twitter           &7.87   	&\textbf{2.52}     &36.90   	&- 		&x 		&379.16	\\   
    \bottomrule

    \end{tabular}

    \caption{SSSP runtime (in seconds) using 48 threads for all engines. ``-'' indicates the engine does not process over 70 million edges. \EHSMALL denotes \EH and the other engines include Galois (G), PowerGraph (PG), a commercial graph tool (CGT-X), and SociaLite (SL). ``x'' indicates the engine did not compute the query properly.}

    \label{table:sssp}
  \end{center}
\end{table}

\subsection{Micro-Benchmarking Results}

\label{sec:eh_optimizations}

We detail the effect of our contributions on query performance. We introduce two new queries and revisit the Barbell query (introduced in \Cref{sec:qc}) in this section: (1) $K_4$ is a 4-clique query representing a more complex graph pattern, (2) $L_{3,1}$ is the Lollipop query that finds all 3-cliques (triangles) with a path of length one off of one vertex, and (3) $B_{3,1}$ the Barbell query that finds all 3-cliques (triangles) connected by a path of length one. We demonstrate how using GHDs in the query compiler and the set layouts in the execution engine can have a three orders of magnitude performance impact on the $K_4$, $L_{3,1}$, and $B_{3,1}$ queries.

\paragraph*{Experimental Setup} These queries represents pattern queries that would require significant effort to implement in low-level graph analytics engines. For example, the simpler triangle counting implementation is 138 lines of code in Snap-R and 402 lines of code in PowerGraph. In contrast, each query is one line of code in \EH. As such, we do not benchmark the low-level engines on these complex pattern queries. We run \texttt{COUNT(*)} aggregate queries in this section to test the full effect of GHDs on queries with the potential for early aggregation. 
The $K_4$ query is symmetric and therefore runs on the same pruned datasets as those used in the triangle counting query in \Cref{sec:triangle}. The $B_{3,1}$ and $L_{3,1}$ queries run on the undirected versions of these datasets.

\begin{table}
  \small
  \centering
     \setlength{\tabcolsep}{2pt}

  \begin{tabular}{@{}llrrrr|rr@{}}
    \toprule
    \bigstrut
    Dataset & Query & \EHSMALL      & -R        & -RA     & -GHD      	  &SL 	& LB \\
    \hline
    \bigstrut
    \multirow{3}{*}{Google+}
    & $K_4$       &4.12       &10.01x       &10.01x       &-          &t/o	&t/o \\ 
    & $L_{3,1}$   &3.11       &1.05x        &1.10x        &8.93x      &t/o &t/o\\ 
    & $B_{3,1}$   &3.17       &1.05x        &1.14x        &t/o        &t/o &t/o\\ 
    \hline
    \bigstrut
    \multirow{3}{*}{Higgs}
    & $K_4$       &0.66       &3.10x        &10.69x       &-          &666x &50.88x\\ 
    & $L_{3,1}$   &0.93       &1.97x        &7.78x        &1.28x      &t/o &t/o \\ 
    & $B_{3,1}$   &0.95       &2.53         &11.79x       &t/o        &t/o &t/o \\ 
    \hline
    \bigstrut
    \multirow{3}{*}{LiveJournal}
    & $K_4$       &2.40      &36.94x        &183.15x      &-          &t/o &141.13x \\ 
    & $L_{3,1}$   &1.64      &45.30x        &176.14x      &1.26x      &t/o &t/o \\ 
    & $B_{3,1}$   &1.67      &88.03x        &344.90x      &t/o        &t/o &t/o \\ 
    \hline
    \bigstrut
    \multirow{3}{*}{Orkut}
    & $K_4$       &7.65     &8.09x          &162.13x      &-          &t/o 	&49.76x \\ 
    & $L_{3,1}$   &8.79     &2.52x          &24.67x       &1.09x      &t/o		&t/o \\ 
    & $B_{3,1}$   &8.87     &3.99x          &47.81x       &t/o        &t/o		&t/o \\ 
    \hline
    \bigstrut
    \multirow{3}{*}{Patents}
    & $K_4$       &0.25     &328.77x        &1021.77x     &-          &20.05x	   &21.77x \\ 
    & $L_{3,1}$   &0.46     &104.42x        &575.83x      &0.99x      &318x   &62.23x \\ 
    & $B_{3,1}$   &0.48     &200.72x        &1105.73x     &t/o        &t/o   &t/o \\ 
    \bottomrule
    \bigstrut
 \end{tabular}
 \caption{4-Clique ($K_4$), Lollipop ($L_{3,1}$), and Barbell ($B_{3,1}$) runtime in seconds for \EH (\EHSMALL) and relative runtime for SociaLite (SL), LogicBlox (LB) and \EH while disabling features. ``t/o'' indicates the engine ran for over 30 minutes. ``-R'' is \EHSMALL without layout optimizations. ``-RA'' is \EHSMALL without both layout (density skew) and intersection algorithm (cardinality skew) optimizations. ``-GHD'' is \EHSMALL without GHD optimizations (single-node GHD).}
 \label{table:advanced_queries}
\end{table}

\subsubsection{Query Compiler Optimizations}

GHDs enable complex queries to run efficiently in \EH{}. \Cref{table:advanced_queries} demonstrates that when the GHD optimizations are disabled (``-GHD''), meaning a single node GHD query plan is run, we observe up to an 8x slowdown on the $L_{3,1}$ query and over a three orders of magnitude performance improvement on the $B_{3,1}$ query. Interestingly, density skew matters again here, and for the dataset with the largest amount of density skew, Google+, \EH observes the largest performance gain. GHDs enable early aggregation here and thus eliminate a large amount of computation on the datasets with large output cardinalities (high density skew). LogicBlox, which currently uses only the generic worst-case optimal join algorithm (no GHD optimizations) in their query compiler, is unable to complete the Lollipop or Barbell queries across the datasets that we tested. GHD optimizations do not matter on the $K_4$ query as the optimal query plan is a single node GHD.

\subsubsection{Execution Engine Optimizations}

\Cref{table:advanced_queries} shows the relative time to complete graph
queries with features of our engine disabled.   
The ``-R'' column represents \EH{} without SIMD set layout optimizations and therefore density skew optimizations. This most closely resembles the implementation of the low-level engines in
\Cref{table:tcount}, who do not consider mixing SIMD friendly layouts. \Cref{table:advanced_queries} shows that our set layout optimizations consistently have a two orders of magnitude performance impact on advanced graph queries. The ``-RA'' column shows \EH{} without density skew (SIMD layout choices) and cardinality skew (SIMD set intersection algorithm choices). Our layout and algorithm optimizations provide the largest performance advantage (>20x) on extremely dense (\bitset) and extremely sparse (\uint) set intersections \cite{DBLP:journals/corr/AbergerNOR15} , which is what happens on the datasets with low density skew here. Like others \cite{Lemke:2010:SUQ:1881923.1881936}, we found that explicitly disabling SIMD vectorization, in addition to our layout and algorithm choices, decreases our performance by another 2x (see \Cref{sec:pruning}). Our contribution here is the mixing of data representations (``-R'') and set intersection algorithms (``-RA''), both of which are deeply intertwined with SIMD parallelism. In total, \Cref{table:advanced_queries} and our discussion validate that the set layout and algorithmic features have merit and enable \EH to compete with graph engines.

\section{Related work}

\label{sec:relatedwork}

Our work extends previous work in four main areas: join processing, graph
processing, SIMD processing, and set intersection processing.

\paragraph*{Join Processing} The first worst-case optimal join
algorithm was recently derived~\cite{worst}. The LogicBlox (LB)
engine~\cite{veldhuizen2012leapfrog} is the first commercial database
engine to use a worst-case optimal algorithm. 
Researchers have also investigated worst-case optimal joins in
distributed settings \cite{chu2015theory} and have looked at
minimizing communication costs \cite{ilprints1102} or processing on compressed
representations~\cite{olteanu2015size}.  Recent theoretical advances
\cite{khamis2015sf,joglekar2015aggregations} have suggested worst-case
optimal join processing is applicable beyond standard join pattern
queries. We continue in this line of work. The algorithm in \EH is a
derived from the worst-case optimal join algorithm~\cite{worst} and
uses set intersection operations optimized for SIMD
parallelism, an approach we exploit for the first time.  Additionally,
our algorithm satisfies a stronger optimality property that we
describe in \Cref{sec:qc}.

\paragraph*{Graph Processing} 
Due to the increase in main memory sizes, 
there is a trend toward developing shared memory graph analytics engines.  
Researchers have released high performance shared memory
graph processing engines, most notably SociaLite
\cite{seo2013socialite}, Green-Marl
\cite{Hong:2012:GDE:2150976.2151013}, Ligra
\cite{Shun:2013:LLG:2517327.2442530}, and Galois \cite{Nguyen:2013:LIG:2517349.2522739}.
With the exception of SociaLite, each of these engines proposes a new
domain-specific language for graph analytics. SociaLite, based on
datalog, presents a engine that more closely resembles a relational
model.  Other engines such as PowerGraph
\cite{Gonzalez:2012:PDG:2387880.2387883}, Graph-X
\cite{Xin:2013:GRD:2484425.2484427}, and Pregel
\cite{Malewicz:2010:PSL:1807167.1807184} are aimed at scale-out performance. The merit of these specialized approaches against
traditional online analytical processing (OLAP) engines is a source of much debate
\cite{Welc:2013:GAW:2484425.2484432}, as some researchers believe
general approaches can compete with and outperform these specialized
designs \cite{Xin:2013:GRD:2484425.2484427,mcsherryscalability}.
Recent products, such as SAP HANA, integrate graph accelerators as
part of a OLAP engine \cite{rudolf2013graph}. Others \cite{fancase} have shown 
that relational engines can compete with distributed engines \cite{Gonzalez:2012:PDG:2387880.2387883,Malewicz:2010:PSL:1807167.1807184} in the graph 
domain, but have not targeted shared-memory baselines. We hope our work
contributes to the debate about which portions of the workload can be
accelerated.

\paragraph*{SIMD Processing}  Recent research has focused
on taking advantage of the hardware trend toward increasing SIMD
parallelism. DB2 Blu integrated an accelerator supporting specialized heterogeneous
layouts designed for SIMD parallelism on predicate filters and aggregates
\cite{Raman:2013:DBA:2536222.2536233}.  Our approach is similar in spirit to DB2 Blu, but applied specifically to join processing. Other approaches such as WideTable\cite{li2014widetable} and
BitWeaving \cite{Li:2013:BFS:2463676.2465322} investigated and
proposed several novel ways to leverage SIMD parallelism to speed up
scans in OLAP engines. Furthermore, researchers have looked at
optimizing popular database structures, such as the trie
\cite{huberadapting}, and classic database operations \cite{Zhou:2002:IDO:564691.564709} to leverage SIMD parallelism.  Our work is the first
to consider heterogeneous layouts to leverage SIMD parallelism as a means 
to improve worst-case optimal join processing. 

\paragraph*{Set Intersection Processing}
In recent years there has been interest in SIMD sorted set
intersection techniques
\cite{DBLP:journalscorrLemireBK14,inoue2014faster,Schlegel11fastsorted-set,highlyscalable}.
Techniques such as the SIMDShuffling algorithm \cite{highlyscalable}
break the min property of set intersection but often work well on
graph data, while techniques such as SIMDGalloping
\cite{DBLP:journalscorrLemireBK14} that preserve the min property
rarely work well on graph data.  We experiment with these techniques
and slightly modify our use of them to ensure min property of the set
intersection operation in our engine. We use
this as a means to speed up set intersection, which is the core
operation in our approach to join processing.

\section{Conclusion}

We demonstrate the first general-purpose worst-case optimal join processing
engine that competes with low-level specialized engines on standard graph workloads. 
Our approach provides strong
worst-case running times and can lead to over a three orders
of magnitude performance gain over standard approaches due to our use 
of GHDs.
We perform a detailed study of set layouts to exploit
SIMD parallelism on modern hardware and show
that over a three orders of magnitude performance gain can be
achieved through selecting among algorithmic choices for
set intersection and set layouts at different granularities
of the data. Finally, we show that on popular
graph queries our prototype engine can outperform
specialized graph analytics engines by 4-60x 
and LogicBlox by over three orders of magnitude.
Our study suggests that this type of engine is a first
step toward unifying standard SQL and graph processing engines.

{
\vspace{3mm}
\scriptsize
\noindent\textbf{Acknowledgments}
We thank LogicBlox and SociaLite for helpful conversations and verification of our comparisons.
Andres N{\"o}tzli for his valuable feedback on the paper and extensive discussions on the implementation of the engine. Rohan Puttagunta and Manas Joglekar for their theoretical underpinnings. Peter Bailis for his helpful feedback on this work.  
We gratefully acknowledge the support of the Defense 
Advanced Research Projects Agency (DARPA) XDATA Program 
under No. FA8750-12-2-0335 and DEFT Program under 
No. FA8750-13-2-0039, DARPA's MEMEX program and SIMPLEX 
program, the National Science Foundation (NSF) CAREER 
Award under No. IIS-1353606, the Office of Naval Research 
(ONR) under awards No. N000141210041 and No. N000141310129,
the National Institutes of Health Grant U54EB020405 awarded 
by the National Institute of Biomedical Imaging and 
Bioengineering (NIBIB) through funds provided by the 
trans-NIH Big Data to Knowledge (BD2K, http://www.bd2k.nih.gov)
initiative, the Sloan Research Fellowship, the Moore Foundation, 
American Family Insurance, Google, and Toshiba. Any opinions, 
findings, and conclusions or recommendations expressed in this 
material are those of the authors and do not necessarily 
reflect the views of DARPA, AFRL, NSF, ONR, NIH, or the 
U.S. government. 

\bibliographystyle{plain}
\bibliography{references}
}

\begin{appendix}

\section{Appendix for Section 2}

\subsection{Dictionary Encoding and Node Ordering}
\subsubsection{Node Ordering}
\label{sec:node_ordering}
Because \EH{} maps each node to an integer value, it is natural to consider the
performance implications of these mappings. Node ordering can affect the
performance in two ways: It changes the ranges of the neighborhoods and, for
queries that use symmetry breaking, it affects the number of comparisons needed
to answer the query. In the following, we discuss the impact of node ordering
on triangle counting with and without symmetry breaking.

We explore the impact of node ordering on query performance using triangle
counting query on synthetically generated power law graphs with different power
law exponents. We generate the data using the Snap Random Power-Law graph
generator and vary the Power-Law degree exponents from 1 to 3. The best
ordering can achieve over an order of magnitude better performance than the
worst ordering on symmetrical queries such as triangle counting.

We consider the following orderings:

\begin{description}

  \item[Random] random ordering of vertices. We use this as a baseline
    to measure the impact of the different orderings.

  \item[BFS] labels the nodes in breadth-first order.

  \item[Strong-Runs] first sorts the node by degree and then starting from the
    highest degree node, the algorithm assigns continuous numbers to the
    neighbors of each node. This ordering can be seen as an approximation of
    BFS.

  \item[Degree] this ordering is a simple ordering by descending
    degree which is widely used in existing graph systems.

  \item[Rev-Degree] labels the nodes by ascending degree.

  \item[Shingle] an ordering scheme based on the similarity of neighborhoods
    \cite{Chierichetti:2009:CSN:1557019.1557049}.

\end{description}

\begin{table}
  \scriptsize
  \begin{center}
    \begin{tabular}{@{}r|cc@{}}
    \hline
    \bigstrut
    Ordering & Higgs & LiveJournal \\
    \hline
    \bigstrut[h]
    Shingles  & 1.67 & 9.14 \\
    \TheGame{} & 3.77 & 24.41 \\
    BFS  & 2.42 & 15.80 \\
    Degree  & 1.43 & 9.93 \\
    Reverse Degree  & 1.40 & 8.47 \\ 
    \bigstrut[b]
    Strong Run & 2.69 & 21.67 \\
    \hline
 \end{tabular}
 \caption{Node ordering times in seconds on 
 two popular graph datasets.}
 \label{table:node_ordering_times}
  \end{center}
\end{table}

\begin{figure}
  \centering
  \includegraphics[width=0.65\linewidth]{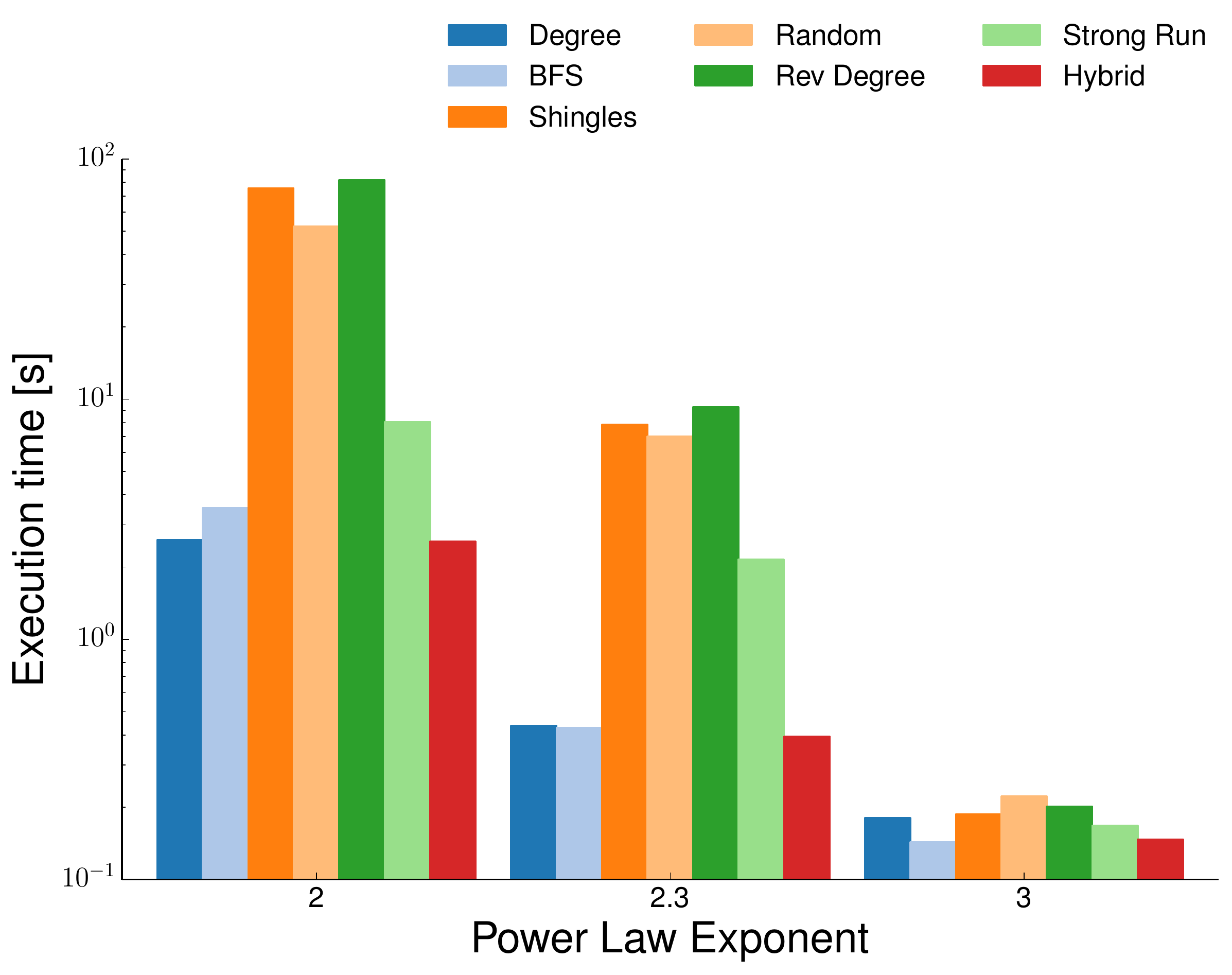}
  \caption{Effect of data ordering on triangle counting with synthetic data.}
  \label{fig:synth_data_ordering}
\end{figure}

In addition to these orderings, we propose a hybrid ordering algorithm
\TheGame{} that first labels nodes using BFS followed by sorting by descending
degree. Nodes with equal degree retain their BFS ordering with respect to each
other. The \TheGame{} ordering is inspired by our findings that ordering by
degree and BFS provided the highest performance on symmetrical queries.
\Cref{fig:synth_data_ordering} shows that graphs with a low power law
coefficient achieve the best performance through ordering by degree and that a BFS
ordering works best on graphs with a high power law coefficient.
\Cref{fig:synth_data_ordering} shows the performance of hybrid ordering
and how it tracks the performance of BFS or degree where each is optimal.

Each ordering incurs the cost of performing the actual ordering of the data.
\Cref{table:node_ordering_times} shows examples of node ordering times in
\EH{}. The execution time of the BFS ordering grows linearly with the number of
edges, while sorting by degree or reverse degree depends on the number of
nodes. The cost of the \TheGame{} ordering is the sum of the costs of the BFS
ordering and ordering by degree.

\subsubsection{Pruning Symmetric Queries}
\label{sec:pruning}

\begin{table}
  \scriptsize
  \centering
  \begin{tabular}{@{}lrr||rr@{}}
    \hline
    \bigstrut
    &  \multicolumn{2}{c||}{Default} & \multicolumn{2}{c}{Symmetrically Filtered} \\
    \cline{2-3}
    \cline{4-5}
    \bigstrut
    Dataset & \uint & \EH{} & \uint & \EH{} \\
    \hline
    \bigstrut[t]
    Google+          & 1.0x & 1.4x & 1.8x & 4.7x \\
    Higgs       & 0.9x & 1.2x & 3.0x & 1.9x \\
    LiveJournal & 1.2x & 1.1x & 1.7x & 1.6x \\
    Orkut       & 1.1x & 1.1x & 1.4x & 1.5x \\
    \bigstrut[b]
    Patents     & 1.2x & 1.1x & 1.9x & 1.3x \\
    \hline
 \end{tabular}
 \caption{Relative time of random ordering compared to ordering by degree.}
 \label{table:node_ordering}
\end{table}

\begin{table}
  \scriptsize
  \centering
  \begin{tabular}{@{}lrrr||rrr@{}}
    \hline
    \bigstrut
    & \multicolumn{3}{c||}{Default} & \multicolumn{3}{c}{Symmetrically Filtered} \\
    \cline{2-4}
    \cline{5-7}
    \bigstrut
    Dataset & -S & -R & -SR & -S & -R & -SR \\
    \hline
    \bigstrut[h]
    Google+ &     1.0x & 3.0x & 7.5x & 1.0x &  4.9x &  13.4x \\
    Higgs &       1.5x & 3.9x & 4.8x & 1.2x &  0.9x &   1.7x \\
    LiveJournal & 1.6x & 1.0x & 1.6x & 1.2x &  0.9x &   1.2x \\
    Orkut &       1.8x & 1.1x & 2.0x & 1.4x &  1.0x &   1.6x \\
    \bigstrut[b]
    Patents &     1.3x & 0.9x & 1.1x & 1.0x &  0.7x &   0.8x \\
    \hline
 \end{tabular}
 \caption{Relative time when disabling features on the triangle counting query. Symmetrically filtered refers to the data preprocessing step which is specific to symmetric queries. ``-S'' is \EH{} without SIMD. ``-R'' is \EH{} using \uint at the graph level.}
 \label{table:breakdown2}
\end{table}

We explore the effect of node ordering on query performance with and without
the data pruning that symmetrical queries enable.  Symmetric queries such as 
the triangle query or the 4-clique query on undirected graphs produce equivalent
results for graphs where each $src,dst$ pair occurs only once and datasets where
each $src,dst$ has a corresponding $dst,src$ pair (the latter producing a result that is
a multiple of the former).  Specialized engines take advantage of restricted optimization that only
holds for symmetric patterns.  For this
experiment, we measure the effect of the node orderings introduced in
\Cref{sec:node_ordering} on five datasets with different set
layouts. We show that node ordering only has a substantial impact on queries
that enable symmetry breaking and that our layout optimizations
typically have a larger impact on the queries which do not enable 
symmetry breaking, which is the more general case.

We use the relative triangle counting performance on 5
datasets with a random ordering and ordering by degree as a proxy for the impact
of node ordering. For each dataset, we measure the triangle counting
performance with random ordering and ordering by degree (the default standard), with and without
pruning, and with the \EH{} set level optimizer and with a homogeneous \uint layout.
We call pruned data on symmetrical queries \emph{symmetrically filtered}. We
report the relative performance of the random ordering compared to ordering by
degree. \Cref{table:node_ordering} 
shows that ordering does not have a large impact on queries that do 
not enable symmetry breaking. In addition,
\Cref{table:node_ordering} shows that our optimizer is more robust to various 
orderings in the special cases where symmetry filtering is allowed. \Cref{table:breakdown2} 
shows that our optimizations typically have a larger impact on data which is not 
symmetrically filtered. This is important as symmetrical queries are infrequent and their symmetrical
property breaks with even a simple selection. 

\subsection{Extended Query Language Discussion}
\label{sec:ext_query_language}
\paragraph*{Conjunctive Queries: Joins, Projections, Selections}
Equality joins are expressed in \EH{} as simple conjunctive queries. We show \EH{}'s' syntax for two cyclic join queries in \Cref{lst:eh_query_language}: the 3-clique query (also known as triangle or $K_3$), and the Barbell query (two 3-cliques connected by a path of length 1). \EH{} easily enables selections and projections in its query language as well. We enable projections through the user directly annotating which attributes appear in the head. We enable selections by directly annotating predicates on attribute values in the body (e.g. b = `Chris').

We illustrate how our query language works by example for the PageRank query:
\begin{example}
\Cref{lst:eh_query_language} shows an example of the syntax used to express the PageRank query in \EH{}. The first line specifies that we aggregate over all the edges in the graph and count the number of source nodes assuming our $Edge$ relation is two-attribute relation filled with $(src,dst)$ pairs. For an undirected graph this simply counts the number of nodes in the graph and assigns it to the relation $N$ which is really just a scalar integer. By definition the $COUNT$ aggregation and by default the $SUM$ use an initialization value of $1$ if the relation is not annotated. The second line of the query defines the base case for recursion. Here we simply project away the $z$ attributes and assign an annotation value of 1/$N$ (where $N$ is our scalar relation holding the number of nodes). Finally, the third line defines the recursive rule which joins the $Edge$ and $InvDegree$ relations inside the database with the new $PageRank$ relation. We $SUM$ over the $z$ attribute in all of these relations. When aggregated attributes are joined with each other their annotation values are multiplied by default \cite{joglekar2015aggregations}. Therefore we are performing a matrix-vector multiplication. After the aggregation the corresponding expression for the annotation $y$ is applied to each aggregated value. This is run for a fixed number (5) iterations as specified in the head.
\end{example}

\section{Appendix for Section 3}

\subsection{Selections}
\label{sec:selections}
\begin{table*}
  \small
  \centering
  \begin{tabular}{ll}
  \toprule
  Name & Query Syntax \\
  \midrule
  4-Clique-Selection &
   \begin{lstlisting}[
    language=C,
    basicstyle=\ttfamily,
    keywordstyle=\bfseries,
    showstringspaces=false]
S4Clique(x,y,z,w) :- R(x,y),S(y,z),T(x,z),U(x,w),V(y,w),Q(z,w),P(x,`node').
  \end{lstlisting} \\
  Barbell-Selection & 
  \begin{lstlisting}[
    language=C,
    basicstyle=\ttfamily,
    keywordstyle=\bfseries,
    showstringspaces=false]
SBarbell(x,y,z,x',y',z') :- R(x,y),S(y,z),T(x,z),U(x,`node'), 
      				       V(`node',x'),R'(x',y'),S'(y',z'),T'(x',z').
  \end{lstlisting} \\
  \bottomrule
  \end{tabular}
  \caption{Selection Queries in \EH}
  \label{lst:eh_sel_query_language}
\end{table*}

Implementing high performance selections in \EH{} requires three additional optimizations that significantly effect performance: (1) pushing down selections within the worst-case optimal join algorithm, (2) index layout trade-offs, and (3) pushing down selections across GHD nodes. The first two points are trivial so we briefly overview them next while providing a detailed description and experiment for pushing down selections across GHDs in \Cref{sec:sel_ghd}. We narrow our scope in this section to only equality selections, but our techniques are general and can be applied to general selection constraints.

\paragraph*{Within a Node} Pushing down selections within a GHD node is akin to rearranging the attribute ordering for the generic worst-case optimal algorithm. Simply put, the attributes with selections should come first in the attribute ordering forcing the attributes with selections to be processed first in \Cref{fig:worst_case}.

\paragraph*{Index Layouts} The data layouts matter again here as placing the selected attributes first in \Cref{fig:worst_case}, causes these attributes to appear in the first levels of the trie which are often dense and therefore best represented as a bitset. For equality selections this is enables us to perform the actual selection in constant time versus a binary search in an unsigned integer array.

\begin{figure}
\begin{subfigure}[b]{0.2\textwidth} 
  \centering
  \includegraphics[height=4cm,width=2.5cm]{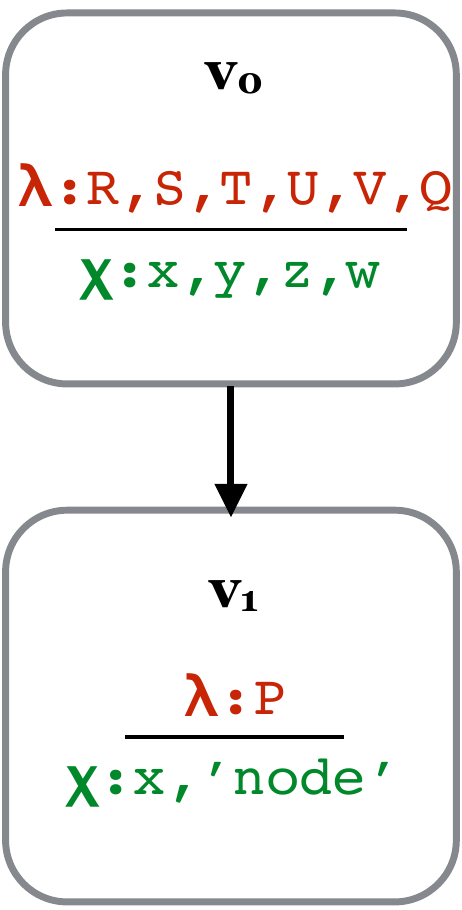}
  \caption{GHD without pushing down}
  \label{fig:4clique}
\end{subfigure}
\begin{subfigure}[b]{0.2\textwidth} 
  \centering
  \includegraphics[height=4cm,width=2.5cm]{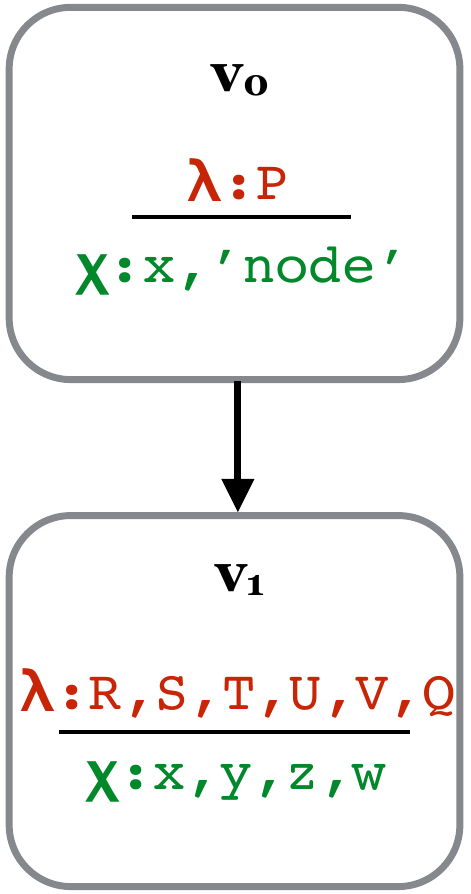}
  \caption{GHD with pushing down}
  \label{fig:np_4clique}
\end{subfigure}
  \centering 
  \caption{We show two possible GHDs for the 4-clique selection query.} \label{fig:ghd_sel}
\end{figure}

\subsubsection{Across Nodes}
\label{sec:sel_ghd}
Pushing down selections across nodes in \EH's query plans corresponds to 
changing the criteria for choosing a GHD described in \Cref{sec:query_plans}. Our goal is to have high-selectivity or low-cardinality nodes be pushed down as far as possible in the GHD so that they are executed earlier in our bottom-up pass. We accomplish this by adding three additional steps to our GHD optimizer:

\begin{enumerate}
\item Find optimal GHDs $\mathcal{T}$ with respect to fhw, changing $V$ in the AGM constraint to be only the attributes without selections.
\item Let $R_s$ be some relations with selections and let $R_t$ be the relations that we plan to place in a subtree. If for each $e \in R_s$, there exists $e' \in R_t$ such that $e'$ covers $e$'s unselected attributes, include $R_s$ in the subtree for $R_t$. This means that we may duplicate some members of $R_s$ to include them in multiple subtrees. 
\item Of the GHDs $\mathcal{T}$, choose a $T \in \mathcal{T}$ with maximal selection depth, where selection depth is the sum of the distances from selections to the root of the GHD. 
\end{enumerate}

\subsubsection{Queries}

To test our implementation of selections in \EH{} we ran two graph pattern queries that contained selections. The first is a 4-clique selection query where we find all 4-cliques connected to a specified node. The second is a barbell selection query where we find all pairs of 3-cliques connected to a specified node. The syntax for each query in \EH is shown in \Cref{lst:eh_sel_query_language}.

Consider the 4-clique selection query: 

\begin{example}

\Cref{fig:ghd_sel} shows two possible GHDs for this query. The GHD on the left is the one produced without using the three steps above to push down selections across GHD nodes. This GHD does not filter out any intermediate results across the potentially high selectivity node containing the selection when results are first passed up the GHD. The GHD on the right uses the three steps above. Here the node with the selection is below all other nodes in the GHD, ensuring that high selectivities are processed early in the query plan. 
\end{example}

\subsubsection{Discussion}

 We run \texttt{COUNT(*)} versions of the queries here again as materializing the output for these queries is prohibitively expensive. We did materialize the output for these queries on a couple datasets and noticed our performance gap with the competitors was still the same. We varied the selectivity for each query by changing the degree of the node we selected. We tested this on both high and low degree nodes. 

 The results of our experiments are in \Cref{table:advanced_sel_queries}.  Pushing down selections across GHDs can enable over a four order of magnitude performance improvement on these queries and is essential to enable peak performance. As shown in \Cref{table:advanced_sel_queries} the competitors are closer to \EH when the output cardinality is low but \EH still outperforms the competitors. For example, on the 4-clique selection query on the patents dataset the query contains no output but we still outperform LogicBlox by 3.66x and SociaLite by 5754x. 

\begin{table}
  \small
  \centering
     \setlength{\tabcolsep}{2pt}
  \begin{tabular}{@{}llr|rr|rr@{}}
    \toprule
    \bigstrut
    Dataset & Query &|Out| & \EHSMALL   & -GHD   &SL 	& LB \\
    \hline
    \bigstrut
    \multirow{4}{*}{Google+}
 	  & \multirow{2}{*}{$SK_4$}       
 		&1.5E+11   &154.24   &6.09x     &t/o	   &t/o   \\ %
    &   &5.5E+7    &1.08     &865.95x   &t/o    &50.91   \\\cline{2-7} %
      & \multirow{2}{*}{$SB_{3,1}$}   
 		&4.0E+17   &0.92     &3.22x     &t/o	   &t/o   \\ %
    &   &2.5E+3    &0.008    &351.72x   &t/o    &t/o   \\\cline{2-7} %
    \hline
    \bigstrut
    \multirow{4}{*}{Higgs}
 	  & \multirow{2}{*}{$SK_4$}       
 		&2.2E+7     &1.92       &14.48x    &t/o	  &58.10x   \\ %
    &   &2.7E+7     &2.91       &9.50x     &t/o    &52.44x   \\\cline{2-7} %
      & \multirow{2}{*}{$SB_{3,1}$}   
 		&1.7E+12    &0.060      &17.36x    &t/o	  &t/o   \\ %
    &   &2.4E+12    &0.070      &14.88x    &t/o    &t/o   \\\cline{2-7} %
    \hline
    \bigstrut
    \multirow{4}{*}{LiveJournal}
 	  & \multirow{2}{*}{$SK_4$}       
 		&1.7E+7     &6.73       &18.05x    &t/o	  &14.83x   \\ %
    &   &5.1E+2     &0.0095     &13E3x &t/o    &10.46x   \\\cline{2-7} %
      & \multirow{2}{*}{$SB_{3,1}$}   
 		&1.6E+12    &0.27       &6.47x     &t/o	  &t/o   \\ %
    &   &9.9E+4     &0.0062     &278.16x   &t/o    &70.23x   \\\cline{2-7} %
    \hline
    \bigstrut
    \multirow{4}{*}{Orkut}
 	  & \multirow{2}{*}{$SK_4$}       
 		&9.8E+8     &208.20     &1.26x     &t/o	  &t/o   \\ %
    &   &2.8E+5     &0.020      &13E+3x &t/o    &18.79x   \\\cline{2-7} %
      & \multirow{2}{*}{$SB_{3,1}$}   
 		&1.1E+15    &3.24       &3.20x     &t/o	  &t/o   \\ %
    &   &2.2E+8     &0.0072     &1314x  &21E+3X    &23E+3x   \\\cline{2-7} %
    \hline
    \bigstrut
    \multirow{4}{*}{Patents}
 	  & \multirow{2}{*}{$SK_4$}       
 		&0         &0.011      &121.70x   &5754x	   &3.66x   \\ %
    &   &9.2E+3     &0.011      &117.56x   &5572x    &10.72x   \\\cline{2-7} %
      & \multirow{2}{*}{$SB_{3,1}$}   
 		&1.6E+1     &0.0060     &77.82x    &223.29x	   &15.17x   \\ %
    &   &1.1E+7     &0.0066     &71.22x    &1073x    &3296x   \\ %
    \bottomrule
 \end{tabular}
 \caption{4-Clique Selection ($SK_4$) and Barbell Selection ($SB_{3,1}$) runtime in seconds for \EH (\EHSMALL) and relative runtime for SociaLite (SL), LogicBlox (LB) and \EH while disabling optimizations. ``|Out|'' indicates the output cardinality. ``t/o'' indicates the engine ran for over 30 minutes. ``-GHD'' is \EH{} without pushing down selections across GHD nodes.}
 \label{table:advanced_sel_queries}
\end{table}

 \subsection{Eliminating Redundant Work}
 
 Our compiler is the first worst-case optimal join optimizer to eliminate redundant work across GHD nodes and across phases of code generation. Our query compiler performs a simple analysis to determine if two GHD nodes
 are identical. 
 For each GHD node in the ``bottom-up'' pass of Yannakakis' algorithm,
 we scan a list of the previously computed GHD nodes to determine if the result
 of the current node has already been computed. We use the conditions below to determine if two GHD nodes are equivalent in the Barbell query. Recognizing this
 provides a 2x performance increase on the Barbell query.

We say that two GHD nodes produce equivalent results in the ``bottom-up pass'' if:
\begin{enumerate}
  \item The two nodes contain identical join patterns on the same input relations.
  \item The two nodes contain identical aggregations, selections, and projections. 
  \item The results from each of their subtrees are identical.
\end{enumerate}

We can also eliminate the ``top-down'' pass of Yannakakis' algorithm if all the attributes appearing in
the result also appear in the root node. This determines 
if the final query result is present after the ``bottom-up'' phase of Yannakakis' algorithm. For example, if we perform a \texttt{COUNT} query on all attributes, the ``top-down'' pass in general is unnecessary. We found 
eliminating the top down pass provided a 10\% performance improvement on the Barbell query.

\section{Appendix for Section 5}

\subsection{Extended Triangle Counting Discussion}
PowerGraph represents each neighborhood using a hash set (with a cuckoo hash) if the degree is larger than 64 and otherwise represents the neighborhood as a vector of sorted node ID's. PowerGraph incurs additional overhead due to its programming model and parallelization infrastructure in a shared memory setting. CGT-X uses a CSR layout and runs Java code for queries which might not be as efficient as native code. Snap-R prunes each neighborhood on the fly using a simple merge sort algorithm and then intersects each neighborhood using a custom scalar intersection over the sets. We note that the runtimes in \Cref{table:tcount} do not reflect the cost of pruning the graph in our system, PowerGraph, SociaLite, or LogicBlox, while CGT-X and Snap-R include this time
in their overall runtime. In Snap-R we found, depending on the skew in the graph, the pruning time accounts for 2\%-46\%
of the runtime on the triangle counting. 

\subsection{Memory Usage}

We utilize a small amount of the available memory (1TB RAM) for the datasets run in this paper. For example, when running the PageRank query on the LiveJournal dataset our engine uses at most 8362MB of memory. For comparison, Galois uses 7915MB and PowerGraph uses 8620MB.

\end{appendix}
\end{document}